\documentclass[aps,prd,twocolumn,notitlepage,
superscriptaddress,
nofootinbib,floatfix]{revtex4-2}

\usepackage[utf8]{inputenc} 
\usepackage{amsfonts,amssymb, mathrsfs, amsmath, esint,bm,enumitem}
\allowdisplaybreaks
\usepackage{siunitx}
\usepackage{bookmark}
\usepackage{multirow}
\usepackage{subcaption}
\usepackage{ragged2e}
\DeclareCaptionJustification{justified}{\justifying}
\makeatletter    
\renewcommand\onecolumngrid{
\do@columngrid{one}{\@ne}%
\def\set@footnotewidth{\onecolumngrid}
\def\footnoterule{\kern-6pt\hrule width 1.5in\kern6pt}%
}
\renewcommand\twocolumngrid{
        \def\footnoterule{
        \dimen@\skip\footins\divide\dimen@\thr@@
        \kern-\dimen@\hrule width.5in\kern\dimen@}
        \do@columngrid{mlt}{\tw@}
}%
\makeatother 

\usepackage{diagbox}
\usepackage{latexsym}
\usepackage{graphicx}
\usepackage{booktabs}
\usepackage[dvipsnames,table]{xcolor}
\usepackage{booktabs,multirow}
\usepackage[capitalize]{cleveref}
\usepackage{titlesec} 
\def\thesection{\arabic{section}}
\def\thesubsection{\arabic{section}.\arabic{subsection}}
\def\thesubsubsection{\arabic{section}.\arabic{subsection}.\arabic{subsubsection}}
\usepackage{physics}
\usepackage{comment}
\usepackage[normalem]{ulem}
\graphicspath{{./Figures/}}

\usepackage{braket}

\usepackage{tikz}
\usetikzlibrary{decorations.pathmorphing}
\usetikzlibrary{math}
\usetikzlibrary{positioning,calc}

\usepackage{hyperref}
\hypersetup{colorlinks, citecolor=ForestGreen, linkcolor=BrickRed, urlcolor=RedViolet}

\newcommand{\tabhspace}{\hspace{.5em}''\hspace{.5em}}
\usepackage{etoolbox}
\usepackage{pgf} 
\newcommand*{\minvalSigma}{1}
\newcommand*{\maxvalSigma}{15}
\newcommand{\gs}[1]{%
    \ifdimcomp{#1pt}{>}{\maxvalSigma pt}{%
    \cellcolor{Goldenrod!80}#1%
    }{%
        \ifdimcomp{#1pt}{<}{\minvalSigma pt}{%
        \cellcolor{Green!80}#1}{%
            \pgfmathparse{int(round(100*(ln(#1/\minvalSigma) / ln(\maxvalSigma/\minvalSigma)))))}%
            \xdef\tempa{\pgfmathresult}%
            \cellcolor{Goldenrod!\tempa!Green!80} #1%
    }}%
}
\newcommand*{\minvalCorr}{0.02}
\newcommand*{\maxvalCorr}{0.93}
\newcommand{\gr}[1]{%
    \ifdimcomp{#1pt}{>}{\maxvalCorr pt}{#1}{%
        \ifdimcomp{#1pt}{<}{\minvalCorr pt}{#1}{%
            \pgfmathparse{int(round(100*(#1-\minvalCorr)/(\maxvalCorr-\minvalCorr)))}%
            \xdef\tempa{\pgfmathresult}%
            \cellcolor{Peach!\tempa!Cyan!20} #1%
    }}%
}

\newcommand{\LCDM}{\Lambda\mathrm{CDM}}
\newcommand{\SInu}{\text{SI}\nu}

\newcommand{\MeV}{\,\text{MeV}}

\newcommand{\eV}{\,\text{eV}}
\newcommand{\meV}{\,\text{meV}}
\newcommand{\Mpc}{\,\text{Mpc}}
\newcommand{\hinvMpc}{\,h\, {\rm Mpc}^{-1}}

\newcommand{\vx}{\mathbf x}

\newcommand{\Geff}{G_\mathrm{eff}}
\newcommand{\Neff}{N_\mathrm{eff}}
\newcommand{\DNeff}{\Delta\Neff}
\newcommand{\aem}{\alpha_\text{em}}
\newcommand{\Ok}{\Omega_k}
\newcommand{\Smnu}{\Sigma m_\nu}

\def\parfrac#1#2{{\left(\frac{#1}{#2}\right)}}

\newcommand{\DR}[1]{}

\newcommand{\DRout}[1]{}
\newcommand{\HZ}[1]{}

\newcommand{\HZout}[1]{}
\newcommand{\PZ}[1]{}

\newcommand{\PZout}[1]{}
\newcommand{\PP}[1]{#1}

\begin{document}
\title{
Neutrino masses from large-scale structures: future sensitivity and theory dependence
}

\author{Davide~Racco} 
\email{dracco@phys.ethz.ch}
\affiliation{Institut f\"ur Theoretische Physik, ETH Z\"urich,Wolfgang-Pauli-Str.\ 27, 8093 Z\"urich, Switzerland}
\affiliation{Physik-Institut, Universit\"at Z\"urich, Winterthurerstrasse 190, 8057 Z\"urich, Switzerland}
\affiliation{Stanford Institute for Theoretical Physics, Stanford University, 382 Via Pueblo Mall, Stanford, CA 94305, U.S.A.}
\author{Pierre~Zhang}
\email{pizhang@phys.ethz.ch}
\affiliation{Institut f\"ur Theoretische Physik, ETH Z\"urich,Wolfgang-Pauli-Str.\ 27, 8093 Z\"urich, Switzerland}
\affiliation{Institute for Particle Physics and Astrophysics, ETH Z\"urich, 8093 Z\"urich, Switzerland}
\affiliation{Dipartimento di Fisica “Aldo Pontremoli”, Universit\`a degli Studi di Milano, 20133 Milan, Italy}
\author{Henry~Zheng}
\email{hz5815@stanford.edu}
\affiliation{Stanford Institute for Theoretical Physics, Stanford University, 382 Via Pueblo Mall, Stanford, CA 94305, U.S.A.}

\begin{abstract}
\noindent
In the incoming years, cosmological surveys aim at measuring the sum of neutrino masses $\Sigma m_\nu$, complementing the determination of their mass ordering from laboratory experiments.
In order to assess the full potential of large-scale structures (LSS), we employ state-of-the-art predictions from the effective field theory of LSS (EFTofLSS) at one loop to perform Fisher forecasts on the sensitivity (combining power spectrum and bispectrum) of ongoing and future surveys (DESI, MegaMapper) in combination with CMB measurements (Planck, Litebird and Stage-4). 
We find that the 1$\sigma$ sensitivity on $\Sigma m_\nu$ is expected to be \PP{15 meV} with Planck+DESI, and \PP{7 meV} with S4+MegaMapper, where \PP{$\sim 10\%$ and $30\%$} of the constraints are brought by the one-loop bispectrum respectively.
To understand how robust are these bounds, we explore how they are relaxed when considering extensions to the standard model, dubbed `new physics'.
We find that the shift induced on $\Smnu$ by a $1\sigma$ shift on new physics parameters (we consider extra relativistic species, neutrino self-interactions, curvature or a time-evolving electron mass) could be $\mathcal O(10)$ meV for Planck+DESI, but it will be suppressed down to $\mathcal O(1)$ meV in S4+MegaMapper.
Our study highlights the quantitative impact of including the bispectrum at one loop in the EFTofLSS, and the robustness of the sensitivity to $\Sigma m_\nu$ against potential new physics thanks to the synergy of cosmological probes. 
\end{abstract}

\maketitle




\section{Introduction}
\label{sec:introduction}
In the incoming years, our experimental knowledge on physics beyond the Standard Model (BSM) will be enriched by crucial information about the neutrino sector \cite{Hernandez:2016kgx, Pascoli:2019wpp}.
Measurements of neutrino oscillations from the experiments T2K, NOvA, JUNO, DUNE, will reveal if the neutrino masses respect the Normal Ordering (NO) or Inverted Ordering (IO), possibly within the next $\sim 5$ years \cite{Cabrera:2020ksc}. 
Current data provide inconclusive hints, partly because of a moderate tension on the phase $\delta_\textsc{cp}$ between NOvA and T2K which is alleviated by IO, and otherwise mildly suggest NO when including the Super-Kamiokande measurement of atmospheric neutrinos  \cite{Esteban:2024eli}.
A direct measurement of neutrino masses is more challenging, and the leading measurement of the endpoint of the spectrum of tritium $\beta$-decays is offered by KATRIN, with $m_\beta < 0.45\eV$ and an expected future reach down to $m_\beta\lesssim 0.2\eV$ \cite{KATRIN:2021uub, Katrin:2024tvg}.
From a cosmological perspective, neutrinos are a significant component of the energy budget of the Universe ($\sim 40\%$ of the total energy density between $T\sim\MeV$ and eV), with important effects on the Large-Scale Structures (LSS) of the Universe (see e.g.~the book~\cite{Lesgourgues:2013sjj} or the reviews \cite{Lattanzi:2017ubx, ParticleDataGroup:2022pth}).
The current epoch of precision cosmology is poised to ``weigh'' \cite{%
Hu:1997mj
} accurately the neutrino impact on the Cosmic Microwave Background (CMB) and LSS, providing the first measurement of the sum of neutrino masses $\Smnu \equiv \sum_{i=1}^3 m_{\nu_i}$ (while a cosmological measurement of single neutrino masses appears to be challenging \cite{Archidiacono:2020dvx,Scott:2024rwc}).  
Cosmological measurements of neutrino masses are indirect measurements. The strong bounds obtained from cosmology will complement future efforts of direct laboratory measurements.

Current upper limits on $\Smnu$ range from $\sim 0.2\eV$ down to $\sim 0.15 \eV$, depending on the datasets included in the fit and on the cosmological model (see \cite{%
Jimenez:2016ckl, 
Archidiacono:2016lnv, 
Vagnozzi:2017ovm,
Raccanelli:2017kht,Vagnozzi:2018pwo, Giusarma:2018jei,
Palanque-Delabrouille:2015pga, Palanque-Delabrouille:2019iyz, 
Pearson:2013iha, Tanseri:2022zfe,
DiValentino:2015sam, 
RoyChoudhury:2018gay,
Capozzi:2021fjo, DiValentino:2021imh, 
diValentino:2022njd, 
DiValentino:2023fei, 
Forconi:2023akg
} and \cite{ParticleDataGroup:2022pth} for a review).
Forecasts for the sensitivity to $\Smnu$ in LSS surveys were performed in \cite{%
Carbone:2010ik, 
DiDio:2013sea, 
Brinckmann:2018owf,
Chudaykin:2019ock,
Euclid:2024imf
}.
Recently, the preliminary results from Baryon Acoustic Oscillations (BAO) released from the first year of the galaxy survey DESI \cite{DESI:2024mwx}, incorporated also in the recent 1-year full-shape DESI analysis \cite{DESI:2024hhd}, have renewed the attention on the implications of the measurement of $\Smnu$ \cite{Craig:2024tky, Notari:2024rti, Allali:2024aiv,Green:2024xbb, Elbers:2024sha, Bottaro:2024pcb, Naredo-Tuero:2024sgf, Jiang:2024viw, RoyChoudhury:2024wri, Loverde:2024nfi}. 
Although the bounds on $\Smnu$ from cosmology might appear impressive when compared to the reach on $m_\beta$ from direct measurements in tritium $\beta$-decay, 
as the precision of the data will soon increase drastically, the challenge for these indirect measurements lies in providing convincing evidence that those are indeed the effects of neutrinos in cosmology, and not other physical effects. 
For these reasons, it is important to extract the largest amount of information to improve the precision of cosmological measurements of $\Smnu$.
Incoming surveys of the CMB from LiteBird and Stage-4 (CMB-S4) experiments, and of galaxies from the ongoing DESI~\cite{DESI:2024hhd} and Euclid~\cite{Euclid:2023bgs} surveys, and future proposals like MegaMapper \cite{Schlegel:2022vrv}, will provide a wealth of data whose most effective deployment relies on the systematic treatment offered by the Effective Field Theory of LSS (EFTofLSS)~\cite{Baumann:2010tm,Carrasco:2012cv,Cabass:2022avo}.

This paper contributes to the question about the robustness of future measurement of $\Smnu$ from LSS against new physics: \textit{how much is it affected by modifications of the cosmological model, the particle content, or BSM neutrino physics?}
This question has been touched upon in the literature (see e.g.~\cite{Archidiacono:2016lnv, Sprenger:2018tdb, Vagnozzi:2018jhn, Brinckmann:2018owf, diValentino:2022njd, Shao:2024mag}).
In our paper, we perform a combined forecast for various CMB and LSS incoming surveys using the state-of-the-art for the EFTofLSS (power spectrum and bispectrum at one loop in perturbation theory~\cite{Perko:2016puo,DAmico:2022ukl}), and we propose to quantify the sensitivity of $\Smnu$ to new physics through the shift induced at 1$\sigma$ by BSM parameters on $\Smnu$.

We consider the following theoretical modifications of SM$+\LCDM$, to cover different directions for new physics.
We focus on additional neutrino self-interactions, decoupled relativistic species contributing to $\Neff$, a curvature component $\Ok$ and variations of SM parameters as $m_e$.
These choices are motivated by the plausibility of similar BSM effects, and by the impact that they have on cosmological degeneracies with $\Smnu$, such as $H_0$.

This paper is structured as follows. 
We summarise in \Cref{sec:theory_models} the main cosmological effects of SM neutrinos, and the theoretical extensions that we consider.
We review in \Cref{sec:EFT-of-LSS} the EFTofLSS, and briefly discuss the potential impact to the model from the presence of massive neutrinos.
\Cref{sec:analysis} reviews our analysis procedure and our results, which are summarised in \Cref{sec:discussion}.
The appendices contain further details about our analysis: \Cref{sec:analytic_fish} provides an analytical estimate of the sensitivity on $\Smnu$ to cross-check the results of our Fisher forecast,
\Cref{sec:priors} describes the priors of our analysis, and
\Cref{sec:m_nu_i} quantifies the effect of the neutrino mass hierarchy for fixed $\Smnu$.


\section{Neutrino cosmology and impact of new physics}
\label{sec:theory_models}
In this section, we briefly summarize the main effects of neutrino masses on the cosmological observables (CMB and LSS) that we analyse, and we review some key aspects of the model modifications that we consider  for their possible connection to the cosmological measurement of $\Smnu$.

\paragraph{Standard neutrino cosmology with $\Smnu$.}
The main effect of a non-vanishing neutrino mass is to turn a relativistic component at early times to a non-relativistic component at late redshifts $z_i^\textsc{nr}+1 = m_{\nu_i}/(0.53 \meV)$ (i.e.~$\gtrsim 110$ for the heaviest neutrino, well after recombination), thus enhancing the total matter component at late times.
Due to their relativistic nature at early times (and possibly still today for one neutrino eigenstate), neutrinos have had a significant time to free stream after their decoupling from the bath. 
Their comoving free-streaming scale, at a time in matter-domination (MD) 
when they are non-relativistic, is%
\footnote{The fractional energy densities of species $\omega_{i,0}\equiv \Omega_{i,0} h^2$ are evaluated at $z = 0$, with $i = b, c, \nu$ standing respectively for baryons, cold dark matter, and neutrinos. The total matter density today is $\omega_{m,0} = \omega_{b,0} + \omega_{c,0} + \omega_{\nu,0}$.}
$k_{\textsc{fs},i} = 1.3\cdot 10^{-2} \Mpc^{-1} \cdot (m_{\nu_i}/0.06\eV) \sqrt{\omega_{m,0}/0.12}/\sqrt{1+z}$. 
The physical free-streaming length was maximised right after neutrinos became non-relativistic, identifying a free-streaming scale $k_{\textsc{nr},i} \equiv k_{\textsc{fs},i}(z=z_{\textsc{nr},i}) \sim \mathcal O(10^{-2}) \hinvMpc$.
On long cosmological scales, neutrinos behave just as a component of cold dark matter which was relativistic at early times.
On short scales $k>k_\textsc{fs}$, neutrinos cannot cluster, and suppress the growth of structure (compared to a Universe where they would be replaced by cold dark matter).
These effects are controlled by the total neutrino mass $\Smnu$, which fixes their physical density $\omega_{\nu,0} = \Smnu/(93.1 \eV)$.

The CMB is affected by $\Smnu$ only for effects related to late-time physics. Primarily, CMB constrains the angular scale of the sound horizon $\theta_s$, whose denominator is the angular diameter distance to recombination $d_A$: 
for fixed $\omega_b,\omega_c$ (well constrained by CMB), a variation of $\Smnu$ affects $h,\omega_\Lambda$ and the time evolution $H^{-1}(z)$ which is integrated over in $d_A$.
The numerator of $\theta_s$, the physical sound horizon $r_s$, depends on the cosmological history up to the recombination epoch, and is not influenced by $\Smnu$.
Further effects involve the integrated Sachs-Wolfe effect and weak lensing \cite{Lesgourgues:2013sjj, Lattanzi:2017ubx, ParticleDataGroup:2022pth}.

The effects of massive neutrinos on structure formation are significant, as they behave as a hot sub-component of dark matter.
At the level of the linear matter power spectrum, if we fix the main parameters that are well-constrained from CMB $(\omega_b, \omega_c,\tau, \theta_s, A_s, n_s)$, the effect of increasing $\Smnu$ is a suppression that is almost flat in $k$ (up to small oscillations for $k\gtrsim k_\textsc{fs}$).
At low $k$, this suppression is due to the decrease in the growth factor%
\footnote{Increasing $\Smnu$, at fixed angular diameter distance $d_A$ and $\omega_m$, implies smaller $h$ and larger $\Omega_m$. Then $H(a)$ grows faster at earlier times, which suppresses the growth factor $D_+\sim \int H^{-3}$.}%
, and coincidentally matches the suppression of the power spectrum at high $k$ due to free-streaming \cite{Archidiacono:2016lnv, ParticleDataGroup:2022pth}.
The effects on the growth of perturbations are discussed in \cref{sec:EFT-of-LSS}.

\paragraph{Beyond the Standard Model: self-interacting neutrinos.}
After discussing first the cosmological property of neutrinos, the transition from relativistic to cold at late times in MD, we now focus on their second feature: they become collisionless after their decoupling from the SM bath, occurring at the freeze-out of the electroweak interactions (when $T_\gamma\sim 1$ MeV in $\LCDM$).
Starting from that time, they develop an anisotropic stress, that impacts structure formation. Their sound speed becomes the speed of light (as long as they are relativistic), rather than $c/\sqrt 3$. 
This impacts the gravitational pull of wavefronts in neutrino overdensities towards the photon and baryon perturbations, thus affecting baryon acoustic oscillations (BAOs) \cite{Stewart:1972, Peebles:1973,Bashinsky:2003tk, Baumann:2015rya,Pan:2016zla}. 

Both from a cosmological and from a particle-physics perspective, it is interesting to consider how this picture gets modified in presence of non-standard self-interactions of neutrinos (SI$\nu$), on top of the electroweak neutral current.
Such interactions between left-handed neutrinos must arise  after electroweak symmetry breaking (EWSB).
A possible origin is from self-interactions of right-handed neutrinos (mediated to $\nu_{\textsc{l}\,i}$ via mass mixing) which can arise if their mass term arises from spontaneous symmetry breaking of a scalar mediator \cite{Berryman:2022hds}. 
Let us remark that, in models with new physics characterised by cut-off scales in the range MeV-GeV (which is typical for cosmological probes of $\Geff$) one generically loses the separation of scales of see-saw models, which accounts naturally for the smallness of $m_{\nu_i}$ \cite{Friedland:2007vv}. 
An interesting phenomenological implication of SI$\nu$ is that they open up the parameter space where right-handed neutrino might achieve the correct relic abundance as dark matter in the Dodelson-Widrow mechanism, where neutrino oscillations naturally populate the right-handed species \cite{Dodelson:1993je,DeGouvea:2019wpf}.

SI$\nu$ are induced by an effective operator proportional to $\Geff \,\overline \nu \nu \, \overline \nu \nu$
below the EWSB scale (with $\nu$ standing for left-handed neutrinos in 2-component notation, and we suppress the flavour indices of $\Geff$ and $\nu$), and are parameterised by the self-scattering rate per neutrino
\begin{equation}
\label{eq:Geff def}
\Gamma = n_\nu \langle \sigma_{\nu\nu\to\nu\nu}v\rangle \equiv \Geff^2 T_\nu^5
\end{equation}
where $\Geff$ is a dimensionful coefficient. In the SM, the EW currents leads to $\Geff^{\textsc{sm}} \sim G_F=1/(\sqrt{2} v_\textsc{ew}^2) = 1.17\cdot 10^{-11}\MeV^{-2}$.
Such interactions face significant constraints e.g.\;from BBN and meson decays, depending on the flavour structure of the couplings \cite{Mirizzi:2013kva, Archidiacono:2013dua, Saviano:2014esa, Forastieri:2015paa, Forastieri:2017oma, Boser:2019rta,Blinov:2019gcj, DeGouvea:2019wpf, Lyu:2020lps, Hagstotz:2020ukm, Grohs:2020xxd, Das:2020xke, Brinckmann:2020bcn}.
In presence of non-standard neutrino interactions as \cref{eq:Geff def}, neutrinos decouple from the SM bath around $T\sim \MeV$ but only start to free stream at a temperature $T_{\nu\textsc{fs}}$  \cite{Brinckmann:2020bcn}%
\begin{equation}
\label{eq:T nu FS}
\frac{T_{\nu\textsc{fs}}}{T_\text{eq}} =\frac{k_{\nu\textsc{fs}} }{k_\text{eq}} \approx \parfrac{\Geff}{0.11\MeV^{-2}}^{-2/3}
\,, \quad (T_{\nu\textsc{fs}} > T_\text{eq})
\end{equation}
where $k_{\nu\textsc{fs}}$ is the mode crossing the Hubble radius at $T_{\nu\textsc{fs}}$.
The effects of $\Geff$ on the CMB and LSS partially overlap with the effects of other cosmological parameters. 
Analyses of CMB from Planck 
\cite{Lancaster:2017ksf, Kreisch:2019yzn, Brinckmann:2020bcn,RoyChoudhury:2020dmd}
and LSS from BOSS
\cite{He:2023oke, Camarena:2023cku} 
reported that a very large value for $\Geff$ can produce a cosmological fit comparable to $\LCDM$. 
From a cosmological perspective, this hint is not particularly robust as the best fits of Planck and BOSS analyses are not compatible \cite{Camarena:2024zck}. 
On the particle-physics side, the ranges of $\Geff$ favoured by the Planck data (a ``strongly-interacting'' scenario with $\log_{10} \Geff^\textsc{SI}/\MeV^{-2}\sim -1.5$, and a ``moderately-interacting'' one with $\log_{10} \Geff^\textsc{MI}/\MeV^{-2}\sim -4$) are almost completely excluded by laboratory constraints on meson decays and cosmological constraints on the abundance of the mediator in the early universe \cite{Blinov:2019gcj}.
In our paper, we consider the model $\LCDM+\Smnu+\Neff+ \Geff$ with the moderately-interacting priors for $\Geff^\textsc{MI}$ listed in \cref{sec:priors}, which are mostly compatible with other constraints, to illustrate the consequences of including $\Geff$ as a free parameter.
We also let $\Neff$ vary, as discussed in the next paragraph.

\paragraph{Beyond the SM: contribution to $\Neff$ from decoupled relativistic species.}
Neutrinos are a relativistic energy component during nucleosynthesis and recombination, which is parameterised by the effective number of neutrino species $\Neff$, defined through the number of relativistic degrees of freedom at $T\ll  m_e$,
\begin{equation}
\label{eq:Neff def}
g_\star (T\ll m_e)= 2+\frac 78 \cdot 2\cdot \Neff \parfrac{4}{11}^{4/3}\,.
\end{equation}
Any deviation in $\Neff$ from its SM value (3.043, with the last digit being refined in recent calculations including finite-temperature QED contributions and neutrino oscillations \cite{Akita:2020szl, Froustey:2020mcq,Bennett:2020zkv, Cielo:2023bqp}) would signal the existence of extra relativistic species. $\Neff$ represents a powerful probe of dark sectors \cite{Blinov:2020hmc, Dvorkin:2022jyg}, notably axions \cite{Ferreira:2018vjj, DEramo:2018vss,Arias-Aragon:2020shv, Ghosh:2020vti, Ferreira:2020bpb, Dror:2021nyr, Green:2021hjh,Bianchini:2023ubu} or sectors explaining neutrino masses \cite{Chacko:2003dt,Gariazzo:2019gyi}, and also a significant constraint for primordial Gravitational Wave backgrounds \cite{Moore:2021ibq, Franciolini:2023wjm}.
For these reasons, $\Neff$ is among the most motivated extensions that are worth considering in cosmological analyses. 
We set it as a free parameter for the models $\LCDM+\Smnu+\Neff+ \Geff$ and $\LCDM+\Smnu+\Neff$.

\paragraph{Beyond $\LCDM$: spatial curvature $\Ok$.}
Among the predictions of the paradigm of primordial inflation to explain the initial conditions of our Universe, the spatial curvature 
\begin{equation}
\label{eq:Ok def}
\Ok\equiv 1-\Omega_{\mathrm{tot},0}
\end{equation} 
would be exponentially diluted and negligible. 
The scalar perturbations sourced during inflation on scales $k \sim H_0$ would appear to us as curvature perturbations, and would be indistinguishable from a curvature component $\Ok \sim \big(k/(a_0 H_0)\big)^2 \zeta(\vec k)|_{k=H_0}$, so that its amplitude would be a random number with RMS $\sqrt{A_s} \sim \mathcal O(10^{-4})$ \cite{Dodelson:2020}.
A detection of $\Ok$ around percent or per mill level would thus be a challenge to the inflationary paradigm. 
The current sensitivity of the combination of Planck, BAO and BOSS is $\Ok <0.2\%$ at 1$\sigma$ \cite{Chudaykin:2020ghx, Vagnozzi:2020rcz, Glanville:2022xes,Simon:2022csv}.
Besides the general importance of this measurement, it is interesting to consider it in connection to neutrino masses, because of their related effects on late-time expansion history and the ``geometric degeneracy'' \cite{Bond:1997wr, Zaldarriaga:1997ch}.

\paragraph{Modifications of SM fundamental parameters: variation of the electron mass $\delta m_e$.}
Some of the SM fundamental parameters have a significant impact on the cosmic history. 
The physics of CMB is determined by electromagnetic interactions in the $\gamma$-$e^-$-$p^+$ plasma, and is sensitive to the values of $\aem$ and $m_e$. 
In particular, $T_\gamma$ and $a_\star$ at recombination can be modified. This possibility was explored especially in connection to the Hubble tension \cite{Uzan:2010pm, Sekiguchi:2020teg, Lopez-Honorez:2020lno, Chluba:2023xqj, Toda:2024ncp}, for which it can provide a reasonably good fit from an observational viewpoint  \cite{Schoneberg:2021qvd}.
In our paper, we consider variations of $m_e$ as an illustrative case. 
On a theoretical standpoint, variations of the electron mass from $m_{e,i}$ at recombination to $m_{e,0}$ at late times (provided its value today is stable enough to avoid stringent laboratory tests) is motivated by the possible existence of light scalar dilaton-like fields coupling to matter, making some fundamental constants field-dependent.
Scalar fields coupled to $m_e \overline e e$, whose mass lies above $H_0$ by a couple of orders of magnitude, and get Planckian initial values after inflation, would naturally provide a variation of $m_e$ after recombination, and would contribute non-negligibly to $\omega_{m,0}$ \cite{Baryakhtar:2024rky}. 
As a word of caution about these models, variations of $m_e$ of order $\sim\mathcal O(10\%)$ would imply a shift in the electron contribution to the cosmological constant much larger than the ambient energy density at recombination%
\footnote{We thank Michael Geller for this observation.}, although the total variation of vacuum energy would also depend on the dynamics of the other fields responsible for the shift.
We define
\begin{equation}
\label{eq:dme def}
\delta m_e \equiv \frac{m_{e,\text{rec}}}{m_{e,0}}\,,
\end{equation}
and we consider the model $\LCDM +\Smnu +\delta m_e$ as an example of the impact of new physics affecting the values of fundamental parameters.

\section{Neutrinos in the Effective Field Theory of Large-Scale Structure}
\label{sec:EFT-of-LSS}

In this section, after providing a short introduction on the EFTofLSS, we review the impact of the presence of massive neutrinos on observables computed within this framework. 

\paragraph{Galaxies at long distances.} A robust measurements of neutrino mass from the LSS can only be contemplated given an accurate description of the gravitational collapse of late-time objects that positions we ultimately observe on the sky.~\footnote{Although it can be any tracers, from here on, we will refer to them as \emph{galaxies}. }
At sufficiently long distance, whatever fills the Universe, dark matter, baryons, galaxies, and so on, has to satisfy the equivalence principle, also called (extended) Galilean invariance in the Newtonian limit~\cite{Jain:1995kx,Scoccimarro:1995if,Weinberg:2003sw,Peloso:2013zw,Kehagias:2013yd,Creminelli:2013poa,DAmico:2021rdb,Marinucci:2024add}. 
Typical variations in the density and velocity fields over a distance $\sim k^{-1}$ scales as $k/k_{\rm NL} \ll 1$ for $k \ll k_{\rm NL}$ once the fields are smoothed over a length scale $\Lambda \gtrsim k_{\rm NL}^{-1}$. 
Building on these considerations, the density field of galaxies can be written into an expansion in powers of the smoothed fields and spatial gradients~\cite{Senatore:2014eva}. 
At each order in perturbations comes a finite number of terms that \emph{i)} have the correct properties under Galilean transformations and \emph{ii)} stem from the gravitational potential $\Phi$ sourced by the massive components of the Universe for which we solve their smoothed, renormalised, equations of motion: dark matter~\cite{Carrasco:2012cv,Carrasco:2013mua}, baryons~\cite{Lewandowski:2014rca,Lewandowski:2015ziq,Braganca:2020nhv}, neutrinos~\cite{Senatore:2017hyk,deBelsunce:2018xtd}, and so on. 
As an EFT, Wilsonian coefficients appearing in the expansions provide a flexible and general parametrisation of our ignorance on the effects of short-scale, nonlinear physics, at the perturbative scales we aim to describe.  
Schematically, the galaxy density fields $\delta^{\rm gal}$ can be written as a sum over scalar operators $\mathcal{O}_i$ multiplied by (time-dependent) Wilsonian coefficients $b_{i}$, 
\begin{equation}\label{eq:field}
\delta^{\rm gal}(\pmb{k}) = \sum_{i} b_{i} \,\mathcal{O}_i(\pmb{k}) \ .
\end{equation}
At each order in perturbations $n$ the Galilean-invariant operators are constructed from $n$ powers of the tidal tensor $s_{ij} \sim \mathcal{H}^{-1} \partial_i \partial_j \Phi$, spatial gradients, and stochastic fields:  $\mathcal{O}_i^{(n)} \equiv \mathcal{O}^{(n)}_i [s_{ij}, \partial_i/k_{\rm M}, \epsilon]$. 
The expansion in spatial gradients accounts for the fact that galaxies that extend over a region of size $\sim k_{\rm M}^{-1}$ are not point-like~\cite{Senatore:2014eva}. 
Stochastic fields, that we collectively denote as $\epsilon$, consist of all quantities whose correlation functions can be written as an expansion in powers of $\partial_i$ preserving rotational invariance~\cite{Carrasco:2013mua,DAmico:2022ukl}.

As galaxies form over a Hubble time, the EFTofLSS is non-local in time~\cite{Carrasco:2013mua,Senatore:2014eva,Donath:2023sav}. 
This implies that more operators than the naive counting suggested by~\eqref{eq:field} appear once displacing the fields along the fluid trajectory~\cite{Senatore:2014eva,Mirbabayi:2014zca}. 
Besides, there are additional contributions when going to redshift space. 
In particular, new counterterms are added to remove the UV-sensitivity of the local products of fields appearing in the redshift-space expansion~\cite{Senatore:2014vja,Perko:2016puo,DAmico:2022ukl}. 
Finally, bulk displacements, that are $\sim \mathcal{O}(1)$ around the BAO scales, have to be properly resummed to describe faithfully the BAO signal~\cite{Senatore:2014via}. 
Adjusting all Wilsonian coefficients in the fit to the cosmological data, this description of the LSS therefore offers controlled, accurate predictions to the cosmological observables, enabling probes of subtle effects on the clustering such as neutrino mass. 

\paragraph{Observables. } For obvious practical reasons, the data, i.e., galaxy maps, are often compressed into summary statistics such as $N$-point functions (see however e.g.,~\cite{Nguyen:2024yth}). 
While the initial distributions of fluctuations in the early Universe is nearly Gaussian, gravity is universal and therefore couples all scales, sourcing non-linear contributions that vanish at the largest scales, however become increasingly important at short distances
, leading to highly non-linear statistics at late times. 
From the expansions at the field level in \cref{eq:field}, the $N$-point functions can be systematically organised into \emph{loop} expansions. 
The higher the loops, the shorter distances can be accessed, therefore increasing the data volume that can be analysed. 
Until recently, most cosmological studies based on the EFTofLSS were limited to low calculations such as the one-loop power spectrum~\cite{Perko:2016puo} or the tree-level bispectrum~\cite{Scoccimarro:1999ed}. 
The constraints on base cosmological parameters, as well as bounds on neutrino mass, has been significantly tightened thanks to the addition of the one-loop correction to the power spectrum (see e.g.,~\cite{DAmico:2019fhj,Ivanov:2019pdj,Colas:2019ret,Chen:2021wdi,Zhang:2021yna,Simon:2022csv}). 
In contrast, the addition of the tree-level bispectrum, because confined to large scales where the signal-to-noise ratio is low, displayed only mild improvements, at the level of $\sim 5-10\%$ (see e.g.,~\cite{DAmico:2019fhj,Philcox:2021kcw}). 
Recently, Ref.~\cite{DAmico:2022ukl} has derived the density of galaxies in redshift space up to the forth order in fields and accordingly the predictions for the bispectrum at the one loop, unlocking the access to information residing at shorter distances beyond the two point.~\footnote{See also~\cite{Philcox:2022frc} for related 
developments.} 
Together with Ref.~\cite{Anastasiou:2022udy} that enables fast evaluation of loop integrals, stringent measurements of cosmological parameters have been obtained from present galaxy data~\cite{DAmico:2022osl,Spaar:2023his} while forecasts indicate that the gain from the addition of the one-loop correction in the bispectrum will be significantly increased with future surveys~\cite{Braganca:2023pcp}.~\footnote{See also~\cite{Cabass:2022epm} for similar analyses 
with the tree-level bispectrum.} 
In light of these recent developments, we provide in the following a re-assessment of the sensitivity of near-future LSS probes to the neutrino mass, considering both the power spectrum and bispectrum in the EFTofLSS at the one-loop order. 

\paragraph{Neutrinos and clustering.} Before moving on, we make a few comments about the treatment of neutrinos in the EFTofLSS. 
In the presence of massive neutrinos, there are several modifications that take place. 
We neglect baryons in the following discussion as their effects can be straightforwardly included in the EFTofLSS~\cite{Lewandowski:2014rca,Lewandowski:2015ziq,Braganca:2020nhv}. 
The relative density fluctuations for matter is then $\delta_m = (1-f_\nu) \delta_{cb} + f_\nu \delta_\nu$, where $f_\nu \equiv \omega_{\nu,0} / \omega_{m,0}$ is the fraction of non-relativistic matter energy density in the form of neutrinos, about $f_\nu \approx 0.4\%$ for $\Smnu = 60 \meV$.
Here $\delta_i(t,\vx) = \rho_i(t,\vx)/\overline{\rho_i}(t) -1$, where $i = m, cb, \nu$ standing respectively for the total matter, cold dark matter + baryons, and neutrinos. 
Therefore the gravitational potential is sourced by an additional contribution of order $f_\nu$, $\partial^2 \Phi = \frac{3}{2} \mathcal{H}^2 \Omega_m (\delta_{cb} + f_\nu \delta_r)$, where we have defined the relative density $\delta_r = \delta_\nu - \delta_{cb}$. 
Because $\delta_r \approx 0$ for $k \ll k_{\rm FS}$ with adiabatic initial conditions and $\delta_r \approx - \delta_{cb}$ for $k \gg k_{\rm FS}$, the time evolution of linear perturbations becomes scale dependent. 
However, for the $k$-range relevant to observations, the linear galaxy overdensity at late time $t_o$ is well captured by
\begin{equation}
\begin{aligned}
\delta^{\rm gal}_{\rm lin}(\pmb{k}, t_o) & = \int^{t_o} \frac{dt}{\mathcal{H}(t)}  \ c(t, t_o) \partial^2 \Phi(\pmb{k}, t) + \dots \\
& \approx b_1(t_o) (1-f_\nu) \delta_{cb}(\pmb{k}, t_o) + \dots \ \label{eq:dg}
\end{aligned} 
\end{equation}
The approximation going to the second line follows from the fact that most of the support of the integral is taken for times $t < t_o$ when $k_{\rm FS}(t) < k_{\rm FS}(t_o) \lesssim k$, for $k$ falling within the observational range. 
Said differently, within the Hubble time the galaxies form most of the neutrinos do not cluster~\cite{Castorina:2013wga}. 
The approximate, scale-independent, linear bias is defined as $b_1(t_o) = \frac{3}{2} \int^{t_o} dt \,\mathcal H(t) c(t, t_o) \Omega_m(t)$, which is the same definition as in the absence of massive neutrinos. 
This matching removes a spurious scale dependence in $b_1$ from the presence of massive neutrinos if one would be using $\delta^{\rm gal}_{\rm lin} \approx b_1 \delta_m$ instead~\cite{Castorina:2013wga,LoVerde:2014rxa,Raccanelli:2017kht,Vagnozzi:2018pwo}. 

We now turn our attention to the linear time dependence of $\delta_c$ in the presence of massive neutrinos within the relevant regime $k \gg k_{\rm FS}$. 
Once non-relativistic, neutrinos slow down the growth of cold dark matter perturbations according to $\delta_{cb} \propto a^{1-\frac{3}{5}f_\nu}$ \cite{Bond:1980ha, Lesgourgues:2006nd}. 
At leading order in $f_\nu$, this gives a relative correction to $\delta_{cb}$ with respect to the case without massive neutrinos of order%
\footnote{Strictly speaking, this comparison between power spectra for massless and massive neutrinos does not keep fixed $\omega_m,\omega_{cb}$, and thus $a_\text{eq}$ between the two cases. For $\Smnu \lesssim 700 \meV$, it is a good approximation as explained in detail in \cite[sec.~6.1.4]{Lesgourgues:2013sjj}.}
$-\frac{3}{5}f_\nu \log \left(a_0/a_\text{nr}\right) \sim - 3 f_\nu$ for $\Smnu \sim 60 \meV$, as $a_0/a_\text{nr} \approx 120 \, \Smnu/ (60\meV)$. 
Together with~\cref{eq:dg}, the main effect of neutrinos on LSS can thus be summarised by a relative correction of the growth of structure of about $-4f_\nu$ for $k \gg k_{\rm FS}$. 
For the power spectrum and bispectrum, this corresponds respectively to a relative suppression of about $1-8f_\nu$ and $1 - 16f_\nu$ \cite{deBelsunce:2018xtd} with respect to the case without massive neutrinos. 
At nonlinear level, the effect of the full scale dependence in the growth of perturbations has been considered also in the loop contributions to the power spectrum~\cite{Blas:2014hya,Garny:2020ilv,Garny:2022fsh,Aviles:2020cax,Noriega:2022nhf}. 
This subleading scale dependence is of size of a few factors of $f_\nu$ times the size of the loop (see further considerations below), and therefore can be neglected for current and near-future sensitivity.  

\paragraph{Renormalisation. } The presence of massive neutrinos implies in principle to revisit the renormalisation in the EFTofLSS. 
First, we realise that for $k \gg k_{\rm FS}$, the structure of the nonlinear contributions will be essentially the same as in the absence of massive neutrinos. 
At the loop level, neutrinos contribute to an amount of the loop size time $\mathcal{O}(10)$ factors of $f_\nu$ coming from the suppression on the growth of cold dark matter perturbations mentioned above.  
New terms contributing mainly at $k \ll k_{\rm FS}$ can be safely neglected: roughly, at one loop, their size is of the order of $\sim k^2 / k_{\rm NL}^2$ times (powers of) the difference $\Delta P = P_{m} - P_{cb}$, where for $k \gg k_{\rm FS}$, $\Delta P / P_{cb} \sim 0$ while for $k \lesssim k_{\rm FS}$, $\Delta P / P_{cb} \sim 2 f_\nu$. 
Here $P_m$ and $P_{cb}$ are respectively the linear power spectrum of the total matter or cold dark matter + baryons only.
New counterterms for these contributions that we are eventually not including can thus also be safely neglected. 
For a proper treatment of renormalisation in the EFTofLSS in presence of massive neutrinos, see~\cite{Senatore:2017hyk}. 

\paragraph{Redshift-space distortions. }
So far, we have only discussed modifications from the presence of massive neutrinos in real space. 
Going to redshift space, positions of galaxies are displaced by peculiar velocities along the line of sight. 
Does the galaxy velocity field, that usually follows the dark matter one (up to higher-derivative terms), receives sizeable corrections from the presence of massive neutrinos? 
For the same reason mentioned earlier, we can convince ourselves that at linear level, galaxies and cold dark matter are in the same bulk motion most of the time within which galaxies form at the scales we observe. 
Another way to see that is by taking the Lagrangian perspective. 
Because the displacement evolution equation is sourced by the gradient of the gravitational potential, the displacement is related to $\delta_{cb}$. 
In this picture, the redshift-space distortions arise from moving the positions of galaxies in real space with the time derivative of the Lagrangian displacement field (see e.g.,~\cite{Matsubara:2007wj}). 
Therefore, our counting of factors of $f_\nu$ contributing to the suppression at $k \gg k_{\rm FS}$ we made earlier in real space must also apply equally for the redshift-space terms. 
Additionally, massive neutrinos generate in the time derivatives of the growing mode of cold dark matter an additional relative suppression of about $\frac{3}{5}  f_\nu$ for $k \gg k_{\rm FS}$ compared to the case where all the matter would be constituted of cold dark matter only, i.e., the growth rate $f(k \gg k_{\rm FS}) \approx (1 - \frac{3}{5} f_\nu)f_0$, where $f_0$ is the (scale-independent) growth rate for $\sum m_\nu = 0$~\cite{Hu:1997vi,Aviles:2020cax,Noriega:2022nhf}. 

\paragraph{Bias expansion. }
To end, we ask ourself if the bias expansion~\eqref{eq:field} is modified in the presence of massive neutrinos. 
There is a fraction of neutrinos slow enough, $v \lesssim v_{\rm NL}$ where $v_{\rm NL} \sim \mathcal{H}/k_{\rm NL}$, that behave effectively as a second matter fluid. 
Thus, in principle, additional contributions in $\delta^{\rm gal}$ arise from the relative density $\delta_r = \delta_\nu - \delta_{cb}$ and velocity $\pmb{v}_r = \pmb{v}_\nu - \pmb{v}_{cb}$ between cold dark matter and neutrinos. 
For non-zero initial relative velocities, the galaxy density receives a contribution from a term $b_{v_r^2} v_r^2$, as it has to be a (Galilean-invariant) scalar. 
For the bispectrum, this correction enters at tree level. 
However, $b_{v_r^2}$ is of the order of $f_\nu^2$ times the size of the initial relative velocity between cold dark matter and neutrinos. 
Moreover, for $k \lesssim k_{\rm NL}$, their relative velocity is vanishing given adiabatic initial conditions. 
Therefore, this correction is negligible.
Let us turn on the term $\propto \delta_r$ that enters at linear level. 
As argued above, $\delta_r$ is vanishing for $k \ll k_{\rm FS}$, and $\sim \delta_{cb}$ for $k \gg k_{\rm FS}$. 
Therefore, for the scales of interest, its net effect is mainly accounted by a rescaling in $b_1$. 
This correction is however suppressed, roughly, by a factor of $f_\nu$ times the fraction of the slow neutrinos that cluster. 
The slow neutrinos are a small fraction, especially once averaged over an Hubble time as we have argued. 
Neglecting it therefore amounts to an error of a small fraction of $f_\nu$ in the amplitude of the power spectrum (and therefore on the determination of the neutrino mass). 
Given that the maximal sensitivity we will find is at most $\mathcal{O}(10) f_\nu $, we neglect this source of error. 
In this context, we conclude that the additional contributions to the bias expansion from the presence of massive neutrinos can be neglected. 
We note that all extensions presented in \cref{sec:theory_models} do not modify the structure of the EFTofLSS predictions, beyond their effects on the linear power spectrum and the growth rate $f$.

\paragraph{Summary of leading $f_\nu$-corrections.} 
We can understand the main sensitivity of LSS on $\Smnu$ by summarising its effect on observables at tree level and leading order in $f_\nu$. 
On scales $k \gg k_{\rm FS}$, which constitute the most relevant range in galaxy surveys where the signal peaks, the amplitudes of the power spectrum and the bispectrum of galaxies are suppressed by relative corrections of about $-8f_\nu$ and $-16f_\nu$, respectively, compared to a Universe where $\Smnu = 0$. 
This stems from two suppressions on the linear galaxy field, $\delta_g$. 
First, there is the well-known relative correction of $-3f_\nu$ in the growth function $D_+$ of matter. 
Next, there is another correction of $-1f_\nu$ when relating the galaxy density to the Laplacian of the gravitational potential sourced by the total matter density, as made explicit in~\cref{eq:dg}. 
This overall suppression in $\delta_g$ applies both to real and redshift space: on the range of observed scales which are under control in the EFTofLSS, the whole \emph{phase-space} distribution of galaxies is fully dictated by the universality of free fall, up to subleading corrections from the two-fluid system. 
This means that \emph{both} the density and the velocity of galaxies at large scales are determined by the gravitational potential sourced by the \emph{total} matter. 
In redshift space, there is finally the well-known additional correction on the growth rate $f$, of about $-3f_\nu/5$ with respect to the $\Smnu =0$ limit.  
On long scales $k \lesssim k_{\rm FS}$, the scale dependence in the growth factor, at linear level, is fully accounted for in our analysis when computing the linear power spectrum up to the relevant redshift of the survey using a Boltzmann numerical solver~\cite{blas2011cosmic}.

\PP{To summarise, we use as input linear power spectrum $(1-f_\nu)^2 P_{cb}$ instead of $P_m$ in the EFT predictions, where $P_{cb}$ is computed by \texttt{CLASS}. 
This is a good approximation%
\footnote{The residual impact of neutrino masses as a scale-dependence of the galaxy bias was analysed in~\cite{Chiang:2018laa,Munoz:2018ajr,Giusarma:2018jei}, showing that it amounts to an effect $\sim 0.2- 0.5 f_\nu$, which roughly implies a relative difference of a few percent on $\Smnu$.}
(at fraction of $f_\nu$) within the range of scales observed in LSS, and during the relevant epochs for galaxy formation, most of the neutrinos are free-streaming~\cite{Castorina:2013wga}. 
} 
This amplitude correction factor of $-2f_\nu$ is important in redshift space as the galaxy velocity is unbiased with respect to the one of matter (up to higher derivative corrections), and therefore this factor is not fully absorbed in $b_1$. 
In fact, it affects directly the sensitivity of the LSS to the neutrino mass as $\Smnu$ is measured essentially thanks to redshift-space distortions (once the primordial amplitude is fixed by CMB) as they allow to break the degeneracy with $b_1$.~\footnote{The relative sensitivity on $\Smnu$ loosens roughly by $\sim 2/8$, compared to the prescription $P_m \rightarrow P_{cb}$.  }
This is discussed in detail in \cref{sec:analytic_fish}, where we provide simple analytic Fisher estimates based on our counting of $f_\nu$-factors to cross-check the results from our realistic Fisher forecasts presented in the next section. 
Overall, we find that they align, provided the caveats that we list.


\section{Analysis}
\label{sec:analysis}

\subsection{Fisher Forecast Pipeline}
\label{sec:fisher}
The Fisher information matrix has been widely used in cosmology to provide forecasts of parameter constraints since~\cite{Tegmark:1997rp}. The Fisher information matrix is an efficient tool to estimate changes in parameter constraints when the theory model or experimental designs are modified. Without having to perform the full Bayesian inference analysis using tools such as MCMC on a cluster, the Fisher matrix provides an order one estimate of parameter covariance under the assumption that the fiducial model does not significantly deviate from the true best fit and the true posteriors are close to Gaussian. In this work, we are interested in studying the robustness of neutrino constraints obtained within $\LCDM$ to compare with those from model extensions that can potentially open up degeneracies of new parameters with the neutrino mass for current and future surveys (DESI and MegaMapper). The Fisher information matrix, $F_{ij}$ is defined as the data average of the Hessian of the log-likelihood function,
\begin{equation}
F_{ij} = -\left< \frac{\partial^2 l}{\partial \theta_i \partial \theta_j} \right> = \frac{\partial T}{\partial \theta_i}C^{-1}\frac{\partial T}{\partial \theta_j}
\label{eq:fisher}
\end{equation}
where $l = \ln L$ and $\theta_i$ are the model parameters, and the second equality is true when the data $x$ is Gaussian distributed around the theory mean $T$ with covariance $C$, such that $-2l = \ln \det C + (x-T)C^{-1}(x-T)$. Furthermore for Gaussian likelihoods, the Fisher information matrix can be interpreted as the inverse covariance matrix of the estimated parameters, i.e. $F^{-1}_{ij} = \langle \Delta \theta_i \Delta \theta_j \rangle - \langle \Delta \theta_i\rangle \langle \Delta \theta_j \rangle$, with $\Delta \theta_i = \theta_i - \theta_{i,0}$, where $\theta_{i,0}$ is the true value of $\theta_{i}$. We adopt the same framework outlined in~\cite{Braganca:2023pcp} to compute the Fisher matrix, the only difference being the additional cosmological parameters corresponding to the various $\LCDM$ extensions we consider.

The theoretical model constructed from the aforementioned bias expansion involves the monopole and quadrupole one-loop power spectrum, as well as the monopole one-loop bispectrum of biased tracers in redshift space within the Effective Field Theory of Large-Scale Structure (EFTofLSS), as derived respectively in~\cite{Perko:2016puo,DAmico:2022ukl}. The computation of the Fisher matrix in Eq.~\eqref{eq:fisher} involves taking derivatives of the theory model. For derivatives of cosmological parameters such as $h$ and $n_s$, we use the finite difference method. Derivatives of the EFT parameters are performed analytically. We use the linear Boltzmann equation solver \texttt{CLASS} package~\cite{blas2011cosmic} to produce the linear power spectrum%
\footnote{For the extensions with self-interacting neutrinos, we use a modified version of \texttt{CLASS} made publicly available \href{https://github.com/PoulinV/class_interacting_neutrinos}{here}.%
} and the code of~\cite{Anastasiou:2022udy} to efficiently evaluate both the power spectrum and bispectrum.

Furthermore, the Fisher is to be evaluated on a fiducial cosmology. For the fiducial model, we adopt the cosmological parameters best-fit by the Planck 2018 results~\cite{Planck:2018vyg}, and the best-fit of EFT parameters from BOSS~\cite{DAmico:2022osl}, appropriately scaled by the redshift for each experiment considered from the effective redshifts of BOSS. The choice of scaling we apply to the EFT parameters is the same scaling of EFT parameters as in~\cite{Braganca:2023pcp}, which extrapolates the fiducial values of these parameters from the BOSS best fit to the redshift bins of other surveys such as DESI or MegaMapper.

Given a fiducial cosmology with parameters $\theta_{\rm fid} \subset \{ \Omega_{m, \rm fid}, h_{\rm fid}, \dots \}$ at a fixed redshift $z$, the power-spectrum Fisher matrix is given by,
\begin{equation}
F^{P}_{ij}(z) = \sum^{k_{\rm max}}_{k}\sum_{\ell, \ell' \in \{0,2\}}\left. \frac{\partial P^{\ell}(k)}{\partial \theta_{i}} C^{\ell \ell' -1}_{PP}\frac{\partial P^{\ell'}(k)}{\partial \theta_{j}}\right|_{\theta = \theta_{\rm fid}}
\end{equation}
where $ C^{\ell \ell' }_{PP}(k)$ is the analytical covariance of the power spectrum and $\ell = 0, 2$ are the monopole and quadrupole, i.e.\ $P^{\ell}(k) = \tfrac{2\ell+1}{2} \int \mathrm d\mu \, P(k,\mu) \mathcal L_{\ell}(\mu)$, where $\mathcal L_{\ell}$ are the Legendre polynomials and $P(k,\mu)$ is the redshift space galaxy power spectrum. In particular, $\mu = k_{z}/k$ is the cosine of the angle between $\pmb k$ and the line-of-sight direction $\hat{z}$.
Similarly for the monopole bispectrum $B^{0}(k_{1},k_{2},k_{3})$ the Fisher matrix is given by,
\begin{equation}
F^{B}_{ij}(z) 
=\sum_{(k_{1},k_{2},k_{3})\in \{\bigtriangleup_{k}\}} \left. \frac{\partial B^0}{\partial \theta_{i}} C^{-1}_{BB}\frac{\partial B^{0}}{\partial \theta_{j}}\right|_{\theta = \theta_{\rm fid}}\, ,
\end{equation}
where $C^{-1}_{BB}(k_1,k_2,k_3)$ is the monopole bispectrum covariance and $\bigtriangleup_{k}$ are the set of triangles $\vec{k}_{1}+\vec{k}_{2}+\vec{k}_{3} = 0$ satisfying $k_{\rm min}<k_{i}< k_{\rm max}$. The analytical covariances of the power spectrum and bispectrum are computed using `FKP' weighting for the power spectrum and the bispectrum estimator \cite{Scoccimarro:1997st, Chan:2016ehg}. 
In this work, we use the Gaussian approximation to compute the covariances neglecting higher-order corrections together with the cross-covariance between the power spectrum and bispectrum. 
As such, we simply add their Fisher matrices when forecasting parameter constraints from their combination.
For the LSS surveys, DESI and MegaMapper, we divide the redshift bins into an effective low-redshift bin and high-redshift bin and compute Fisher matrices for the two effective bins instead of for every redshift bin in the survey. For each effective redshift bin, the effective redshift $z_\text{eff}$, shot noise $\overline n^{-1}_\text{eff}$, and survey volume $V_\text{eff}$ are computed as the weighted average of redshift bins, where the shot noise and survey volume in each redshift bin are weighted by the number of tracers in that bin. We use the same specifications as outlined in Table 2 of~\cite{Braganca:2023pcp} for DESI and Table 3 of~\cite{Braganca:2023pcp} for MegaMapper. The analytical covariance of a given redshift bin depends on the survey volume and shot noise through the power spectrum and bispectrum estimators \cite{Chan:2016ehg}. In general, lower shot noises and larger survey volumes lead to higher precision from the data. The full Fisher matrix of a survey is given by the sum of the Fisher matrices of the low-redshift and high-redshift bins, \PP{together with the prior that we impose on the EFT parameters described in~\cref{sec:priors}. 
In particular, we make use of the perturbativity prior described in~\cite{Braganca:2023pcp}, to consistently restrict the EFT predictions within the physically allowed region in perturbation theory. This prior imposes a cap on the size of one-loop contributions to the maximum allowed theory error defined by a two-loop estimate which is constrained to lie within the data error of the survey.}

The non-linear scale $k^{-1}_{\rm NL}$ and the maximum theory reach $k_{\rm max}$ are estimated using BOSS CMASS as the comparative baseline. 
The non-linear scale of an effective redshift bin is estimated by requiring the integrated linear power spectrum to be equal to that of BOSS CMASS, i.e. $\int^{k_{\rm NL}}_0 P^{\rm survey}_{11}(q,z)q^2\mathrm dq=\int^{k_{\rm NL, CMASS} = 0.7}_0 P^{\rm CMASS}_{11}(q,z=0.57)q^2\mathrm dq$. For EFT parameters originating from velocity fields from the redshift space transformation, the non-linear suppression comes from a different scale $k_{\rm NL, R} \simeq k_{\rm NL}/\sqrt{8}$ as noted in \cite{DAmico:2022osl}. The theory reach $k_{\rm max}$ is defined to be the maximum $k$ at which the amplitude of the two-loop contributions, or the theory error, exceeds the data error. While the exact two-loop contributions are not known, their amplitude can be estimated assuming a power-law universe \cite{Carrasco:2013sva}. The $k_{\rm max}$ of a given effective redshift bin of a survey is then estimated by requiring the integrated theory noise to data noise ratio up to $k_{\rm max}$ to be equal to that of CMASS. The exact equation to solve is given by \cite[Eq.~(2.26)]{Braganca:2023pcp}.
For reference, the experiment specifications, effective volume, shot noise, and $k$-reach, for each redshift bins, are summarized in \cref{tab:probes}.
\begin{table}[h!]\centering
\begin{tabular}{c c c c c ccc c} 
\toprule
& bin & $z_{\text{eff}}$ & $\overline n_{\text{eff}}$ 
& $b_1^{\text{ref}}$ & 
$k^{\text{Tree}}_{\rm max}$ & $k^{1L}_{\rm max}$ & $k_{\text{NL}} $ & $V_{\rm eff}$\\
\midrule
\multirow{2}{*}{DESI} & 1 & 0.84& 8.0 & 1.3 & 0.08 & 0.18 & 0.9 &3.5 \\ 
 & 2 &1.23 & 3.2 & 1.5& 0.09 & 0.23 & 1.3 &5.1\\ 
 \midrule
\multirow{2}{*}{MegaMapper} & 1 & 2.4& 18 & 3.1 & 0.14 & 0.36 & 3.2& 27 \\
 &2 &4.3 & 1.1 & 6.3& 0.28 & 0.76 & 10.1 &24\\
 \bottomrule
\end{tabular}
\caption{Effective survey specifications for DESI and MegaMapper. Bins 1 and 2 refer to low- and high-redshift bins at an effective redshift $z_\text{eff}$, and $\overline n_{\text{eff}}$ (in units of $10^{-4}(\hinvMpc)^3$) is the background galaxy number density entering the derivatives (not the covariance), momenta $k$ are in units of $\hinvMpc$, and $V_{\rm eff}$ is the effective survey volume in units of $h^{-3} \mathrm{Gpc}^{3}$.}
\label{tab:probes} 
\end{table}

The CMB forecasts are computed using \texttt{MontePython}~\cite{Brinckmann:2018cvx} with a fake likelihood generated using the same fiducial parameters as the Fisher, $\theta_{\rm fid}$. To combine CMB with LSS results, we extract the inverse covariance of cosmological parameters obtained from the MCMC chains generated by \texttt{MontePython} and simply add this to the LSS Fisher matrix.~\footnote{When the posteriors from CMB are far from Gaussian as when considering the model extensions scrutinised in this work, this procedure becomes quite inaccurate. 
To remedy to this, we instead produce posteriors from a joint analysis of CMB and fake BAO with a large covariance matrix for the latter. 
This combination efficiently breaks the degeneracies inherent to the CMB, such that the resulting posteriors are close to Gaussian. 
Then, upon combination with LSS we can then remove the Fisher information from the injected fake BAO to recover the CMB+LSS forecasts. 
}
In this work, we neglect the small cross-correlation between LSS and CMB lensing.

As noted in~\cite{Braganca:2023pcp}, there are several observational and systematic effects not captured by the Fisher information matrix, which can lead to significant deviations from the true constraints. 
\PP{As a reminder, under the Cramér-Rao bound the inverse Fisher matrix is the lower bound on the covariance of observed parameters.
Nevertheless, when the real posteriors are close to Gaussian (as it is the case in our analysis when we combine LSS and CMB), this is a good approximation.}
As mentioned earlier, one such systematic effect is the analytical approximation of the data covariance. While the covariance of the power spectrum and bispectrum data can be estimated using analytical expressions as per~\cite{Agarwal:2020lov}, this approach includes only diagonal contributions from the tree-level power spectrum. The omission of off-diagonal contributions and loop-level contributions to the diagonal covariance is estimated to result in constraints that are 25\% to 30\% tighter when validated against BOSS~\cite{Braganca:2023pcp}. Furthermore, our Fisher study does not account for the modeling of the Alcock-Paczynski (AP) effect and the window function, which we refer to as `observational' effects. Consequently, the constraints are further estimated to be tightened by an additional 15\% to 30\%. Overall, we therefore estimate that the Fisher forecast may deviate by $\sim50\%$ when compared to results obtained from a full analysis.

While many priors are possible on $H_{0}$, such as the measurements from the SH0ES collaboration or the H0LiCOW collaboration~\cite{Riess:2019cxk, H0LiCOW:2019pvv}, we choose not to include any such priors in our analysis. For one, this provides a more conservative analysis of the robustness of neutrino physics. Secondly, the existing $H_{0}$ tension makes the choice of a particular prior difficult. Lastly, with the full shape analysis~\cite{Colas:2019ret, DAmico:2020ods}, the galaxy clustering constraining power on $H_{0}$ is sufficient and comparable to the aforementioned priors~\cite{DiValentino:2015sam}.
The specifications for the fiducial model and the priors are detailed in~\cref{sec:priors}.

\subsection{Survey specifications}
We collect here the information about the three future surveys that we consider in our study.
\begin{itemize}
    \item \textbf{S4} includes LiteBIRD and CMB-S4. 
LiteBIRD (Lite (Light) satellite for the studies of B-mode polarization and Inflation from cosmic background Radiation Detection) is a planned space telescope to be launched from the Tanegashima Space Center and will become operational in the next 10 years. LiteBIRD aims to detect primordial density fluctuations and its imprint in the CMB ``B-mode'' polarization~\cite{LiteBIRD:2022cnt}. 
The CMB-S4 is a proposed ground-based, ultra-deep survey covering 40\% of the sky over 7 years, aimed at improving by one order of magnitude the sensitivity compared to Stage-3 CMB experiments~\cite{CMB-S4:2022ght}. 
We note that no recent planning for CMB-S4 have been commissioned, however other incoming experiments like Simons Observatory will reach an equivalent sensitivity~\cite{SimonsObservatory:2018koc}.
For our forecasts, we will complement the low $l$ modes of LiteBIRD $2 \leq l \leq 50$ with CMB-S4 modes $51\leq l\leq 3000$ using their mock likelihoods in the public \texttt{MontePython} package~\cite{CMB-S4:2022ght, Euclid:2024imf, Brinckmann:2018cvx, Audren:2012wb}.
    \item \textbf{DESI} is a recently operational Stage-IV ground-based dark energy experiment aimed at studying the BAO and the spectra of galaxies and quasars. Specifically, DESI targets Bright Galaxy Samples (BGS), Luminous Red Galaxies (LRGs), Emission-line Galaxies (ELGs) and quasars (QSOs)~\cite{DESI:2023bgx, DESI:2016fyo}. We use the same redshift binning, $k_{\rm max}$, linear bias, and shot-noise specifications for DESI as outlined in Table 2 of~\cite{Braganca:2023pcp}. These estimations are based on the DESI survey design proposed in~\cite{DESI:2016fyo}, corresponding to the 5-years plan.
    \item \textbf{MegaMapper} is a proposed ground-based Stage-V dark energy experiment that will observe galaxy samples in a high-redshift range $2 \leq z \leq 5$. The experiment is planned to become operational in around 10 years, following the same timescale as LiteBIRD and CMB-S4 \cite{Schlegel:2022vrv}. 
 We use the same redshift binning, $k_{\rm max}$, linear bias, and shotnoise specifications for DESI as outlined in Table 3 of~\cite{Braganca:2023pcp}. There are two scenarios presented in ~\cite{Schlegel:2022vrv}, ``idealized'' and ``fiducial''. For our forecast, we choose to present our results only for the ``idealized'' scenario, which are based on the specifications in Table 1 of~\cite{Ferraro:2019uce}.
\end{itemize}

\subsection{Neutrino Parametrization}
Usually, when scanning over the neutrino mass with sampling algorithms such as MCMC, it is simple to ensure the positive definiteness of the neutrino mass by using a flat prior with non-negative bounds. However, for a Fisher forecast, it is difficult to introduce flat priors as it violates the Gaussian assumptions. Thus, instead of using a flat prior for the Fisher forecast, we choose to sample instead in log neutrino mass to ensure its positive definiteness. The Gaussian assumption then imposes that the log neutrino posterior is approximately a log normal distribution. To convert 1-$\sigma$ bounds of a log normal distribution to 68\% or 95\% CI's in linear space, we simply use the quantile of a log normal distribution. As such, the 68\% CI of neutrino mass is then given by
\begin{multline}
F^{-1}_{\Smnu}(p = 0.84) - F^{-1}_{\Smnu}(p = 0.16) =\\
=\Smnu^{\rm fid}\cdot\Big(e^{\sqrt{2} \sigma \erf^{-1}(0.68)} - e^{\sqrt{2} \sigma \erf^{-1}(-0.68)} \Big)\, ,
\label{eq:mnuCI}
\end{multline}
where $F^{-1}_{\Smnu}$ is the inverse CDF of the lognormal distribution of $\Smnu$, and $\Smnu^{\rm fid}$ is the fiducial neutrino mass. When combining with Planck and S4 samples, as these samples are sampled in linear space instead of log space, we first re-weight the collected samples by the Jacobian $-\log\left( \Smnu\right)$ before combining with the Fisher matrix of LSS. When Planck or LiteBird + CMB-S4 constraints are presented standalone, the samples are not re-weighted.

\subsection{Results}
\paragraph{Forecast of sensitivity on $\Smnu$ in the baseline model ($\LCDM+\Smnu$ at NO minimal mass), with EFTofLSS including the 1-loop bispectrum.}

\begin{figure}[h!]\centering
\includegraphics[width=.9\columnwidth]{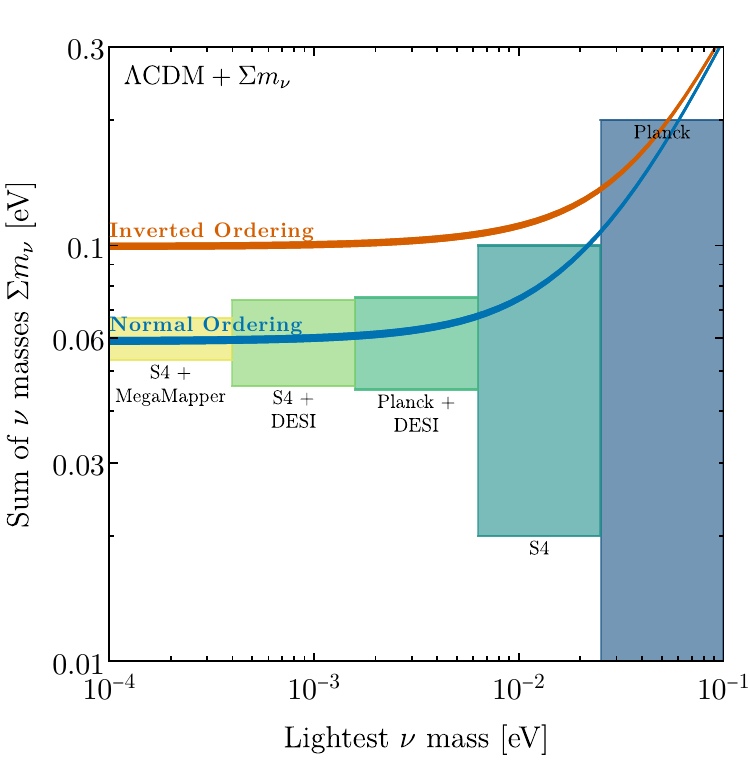}
\caption{Forecast of uncertainty on $\Smnu$ (assuming NO) for various combinations of CMB and LSS experiments.}
\label{fig:mnu_bands}
\end{figure}

The choice of the fiducial minimal neutrino mass $\Smnu$ depends on the choice of the mass ordering. In the NO scenario, the minimal neutrino mass is $\Smnu = 0.06\, \eV$, while for the IO scenario, the minimal neutrino mass is $\Smnu = 0.10\, \eV$. For the main results we will present them with NO minimal mass. To establish a baseline, we present in~\cref{fig:mnu_bands} the neutrino bounds when analyzing the minimal extension $\LCDM + \Smnu$. 
As shown in~\cref{fig:mnu_bands}, the combination of Planck and DESI analysis provides roughly 3 times tighter constraints compared to S4 alone. Furthermore, in the case of S4 + MegaMapper, the constraint is improved to 6 times tighter compared to S4 alone. Comparing the Planck + DESI forecast to S4 + DESI, we see that the gain is mild. However, the forecast of S4 + MegaMapper provides a two times tighter constraint compared to that of S4 + DESI, hence highlighting the significant information gain from LSS data. We will show in the next section that such findings are consistent with all the $\LCDM$ extensions we consider.

\begin{figure}[h!]\centering
\begin{tikzpicture}
\node [above right,inner sep=0] (image) at (0,0) {
\includegraphics[width=\columnwidth, trim=0 0 0 0,clip]{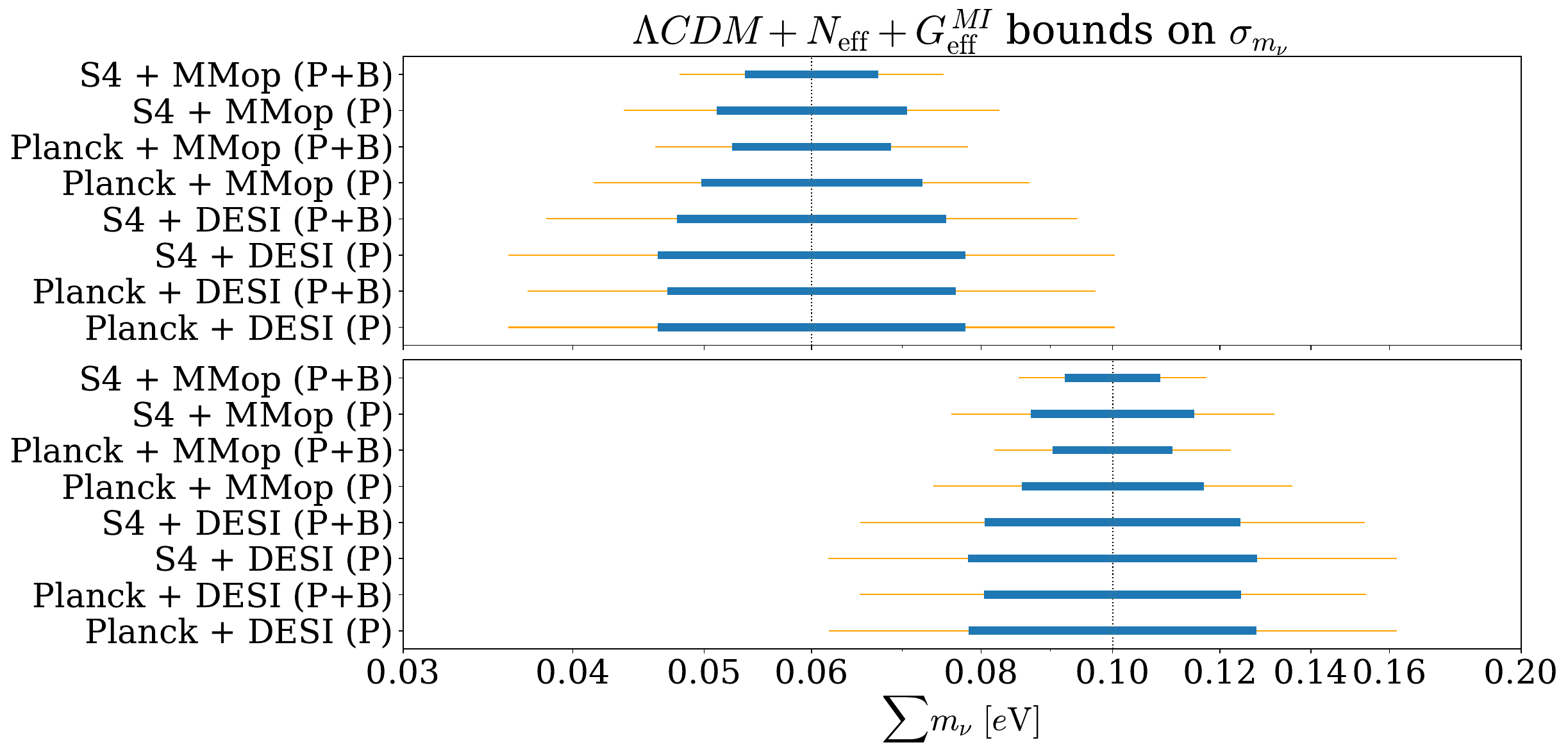}};
\begin{scope}[
x={($0.05*(image.south east)$)},
y={($0.05*(image.north west)$)}]
\filldraw [fill=white,fill opacity=1, draw=white] (0,2.45) rectangle (5,19);
\node[fill=white,align=center] at (10,20.05) {Bounds on $\Smnu$, model $\LCDM+\Neff+\Geff^\textsc{(mi)}$};
\node[fill=white,align=center] at (13.,0.5) {$\Smnu$ [eV]};
\begin{scriptsize}
\node[align=center,text=NavyBlue] at (16.5,14.5) {$\Smnu=60 \meV$\\ \textbf{Normal}\\ \textbf{Ordering}};
\node[align=center,text=RedOrange] at (8,6.5) {$\Smnu=100 \meV$\\ \textbf{Inverted}\\ \textbf{Ordering}};
\node[anchor=west] at (0,17.7) {S4+MegaM.};
\node[anchor=west] at (0,15.7) {Pl.+MegaM.};
\node[anchor=west] at (0,13.7) {S4+DESI};
\node[anchor=west] at (0,11.7) {Pl.+DESI};
\node[anchor=west] at (0,9.3) {S4+MegaM.};
\node[anchor=west] at (0,7.3) {Pl.+MegaM.};
\node[anchor=west] at (0,5.3) {S4+DESI};
\node[anchor=west] at (0,3.3) {Pl.+DESI};
\end{scriptsize}
\begin{tiny}
\node[anchor=east] at (5.2,18.15) {P+B};
\node[anchor=east] at (5.2,17.25) {P};
\node[anchor=east] at (5.2,16.15) {P+B};
\node[anchor=east] at (5.2,15.25) {P};
\node[anchor=east] at (5.2,14.15) {P+B};
\node[anchor=east] at (5.2,13.25) {P};
\node[anchor=east] at (5.2,12.15) {P+B};
\node[anchor=east] at (5.2,11.25) {P};
\node[anchor=east] at (5.2,9.75) {P+B};
\node[anchor=east] at (5.2,8.85) {P};
\node[anchor=east] at (5.2,7.75) {P+B};
\node[anchor=east] at (5.2,6.85) {P};
\node[anchor=east] at (5.2,5.75) {P+B};
\node[anchor=east] at (5.2,4.85) {P};
\node[anchor=east] at (5.2,3.75) {P+B};
\node[anchor=east] at (5.2,2.85) {P};
\end{tiny}
\end{scope} 
\end{tikzpicture}
\vspace*{-2em}
\caption{Summary of 68\% (thick blue) and 95\% (thin orange) CI bounds on $\Smnu$ for the self-interacting neutrino model shown for both analyzing power spectrum only (P) and analyzing power spectrum and bispectrum (P+B). The top panel shows bounds for the minimal $\Smnu$ in NO and the bottom one for IO.}
\label{fig:NeffGeff_all_bounds}
\end{figure}

\paragraph{Small impact of minimal $\Smnu$, and impact of including the 1-loop bispectrum in the EFTofLSS.}
One particularly important theoretical prior is the neutrino mass ordering. We assess how the fiducial mass considered in our Fisher forecast impact the resulting constraints. 
\PP{We stress that here we only want to understand the impact of the fiducial value for $\Smnu$ on the constraints: as it is well known, in the low neutrino mass regime, the effects in cosmology are all well described as if the three neutrino species were degenerate in mass \cite{Archidiacono:2020dvx,Herold:2024nvk}. 
This is also our choice for the analysis of this paper, and we quantify in \cref{sec:m_nu_i} the impact of different choices for the individual neutrino masses.\\
In order to test the negligible dependence of our Fisher forecast on the assumed $\Smnu$, we compare in~\cref{fig:NeffGeff_all_bounds} the results obtained with $\Smnu=60$ or 100 meV. 
For these two cases, we set the individual neutrino masses to the values fixed respectively by normal and inverted ordering. We discuss in \cref{sec:m_nu_i} the impact of the assumed mass hierarchy (NO, IO, degenerate or only 1 massive $\nu$) on LSS measurements.}
\newline
The choice of minimal mass in particular determines the fiducial neutrino mass entering the linear power spectrum used to evaluate the Fisher matrix. 
Both choices lead to similar constraints, and the relative improvements of adding in the bispectrum and using next generation surveys are the same for the minimal mass choices. 
For example, when analyzing Planck + DESI, the bispectrum tightens the neutrino bounds by \PP{8.2\%} in the NO scenario compared to \PP{5.2\%} in the IO scenario. 
The tightest constraint comes unsurprisingly from the combination of S4 + MegaMapper, with \PP{$\sim$ 15\%} improvement upon Planck + MegaMapper. 
We obtain more appreciable$\sim$ 50\% improvements on neutrino constraints when using future LSS experiments as we swap DESI to MegaMapper due to the larger reach in $k$.
We note also that the one-loop bispectrum becomes significant to neutrino constraints when analyzing MegaMapper over DESI, with Megamapper yielding \PP{$\sim$ 49\%} tighter bounds when analyzed with S4 and \PP{$\sim$ 51\%} tighter bounds when analyzed with Planck compared to DESI.

\begin{figure}[h!]\centering%
\begin{tikzpicture}
\node [above right,inner sep=0] (image) at (0,0) {
\includegraphics[width=\columnwidth]{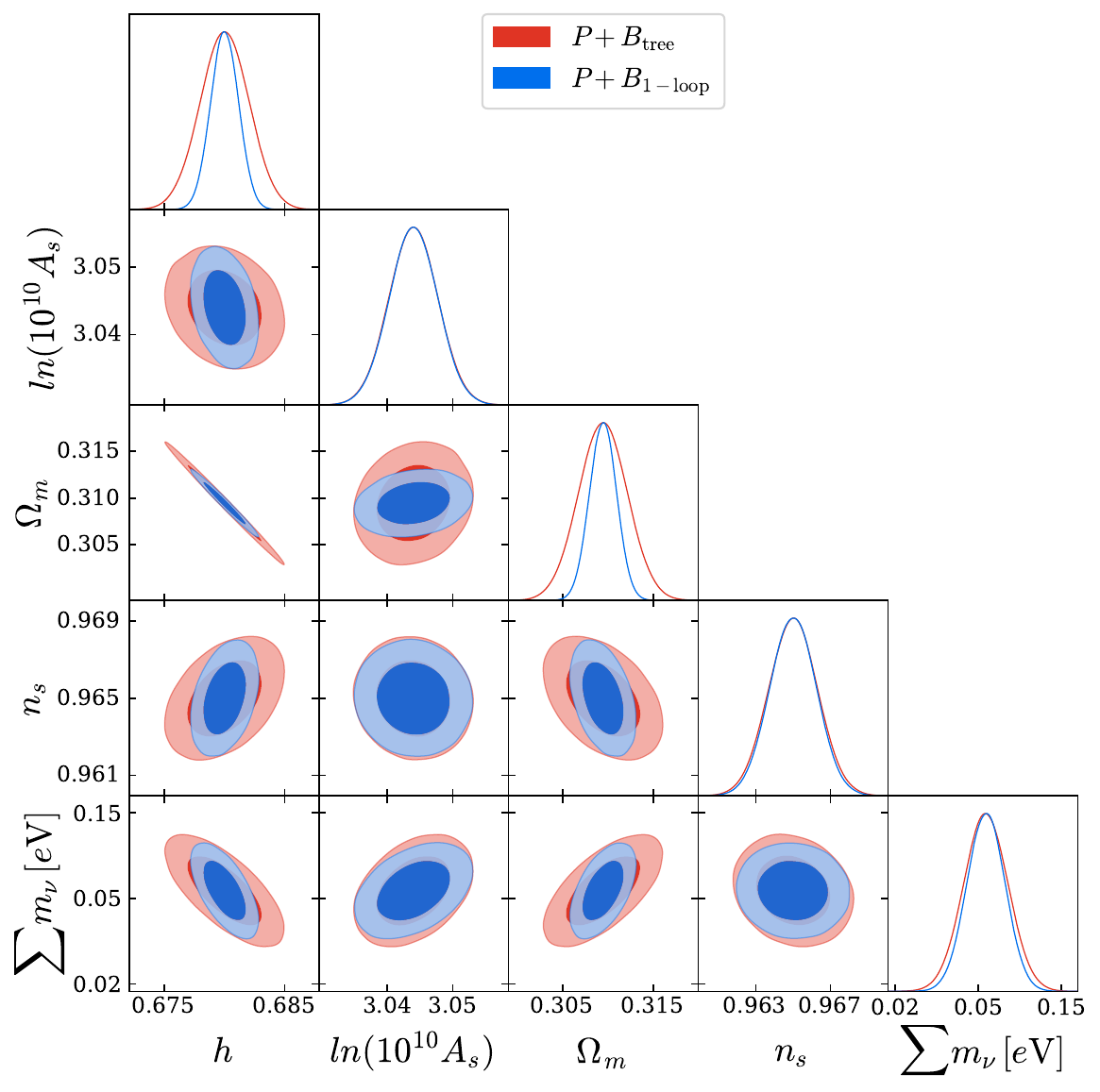} 
};
\begin{scope}[
x={($0.05*(image.south east)$)},
y={($0.05*(image.north west)$)}]
\node[fill=White, rounded corners=3mm, fill opacity=0.5, text opacity=1,align=center] at (17,19) {{\large S4 + DESI}};
\end{scope} 
\end{tikzpicture}
\caption{Constraints on cosmological parameters for 
{S4 + DESI} for tree-level bispectrum vs.\,monopole one-loop bispectrum.}%
\label{fig:S4DESI_btreeloop}%
\end{figure}
\begin{figure}[h!]\centering%
\begin{tikzpicture}
\node [above right,inner sep=0] (image) at (0,0) {
\includegraphics[width=\columnwidth]{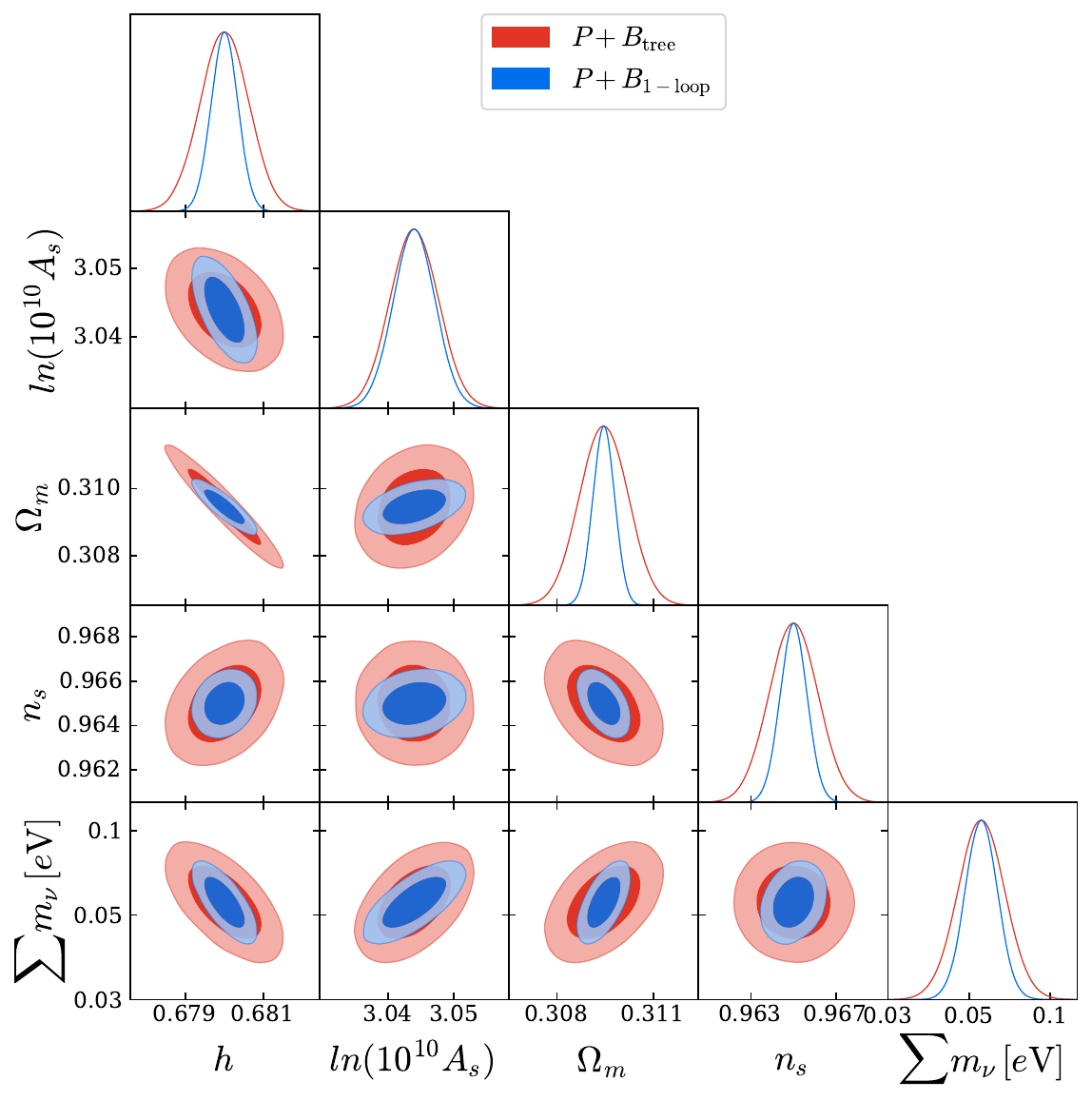} 
};
\begin{scope}[
x={($0.05*(image.south east)$)},
y={($0.05*(image.north west)$)}]
\node[fill=White, rounded corners=3mm, fill opacity=0.5, text opacity=1] at (17,19) {{\large S4 + }};
\node[fill=White, rounded corners=3mm, fill opacity=0.5, text opacity=1] at (17,18) {{\large MegaMapper}};
\end{scope} 
\end{tikzpicture}
\caption{Constraints on cosmological parameters for 
{S4 + MegaMapper} for tree level bispectrum vs.\,monopole one-loop bispectrum.}%
\label{fig:S4MMop_btreeloop}%
\end{figure}
%
The addition of the one-loop bispectrum on top of the tree level yields a \PP{$\sim$15\%} improvement on the constraint of neutrino masses when analyzing S4 + DESI as shown in~\cref{fig:S4DESI_btreeloop}, with more significant improvements also in $h, \Omega_{m}$ and $n_{s}$. This improvement is further extended in the case of S4 + MegaMapper, providing a \PP{$\sim$ 25\%} improvement on the constraint of neutrino masses as shown in~\cref{fig:S4MMop_btreeloop}. Such improvement is largely attributed to the higher reach in $k$ of MegaMapper with $k_{\rm max} \sim 0.9 \unit{h} \, \unit{Mpc}^{-1}$ compared to DESI's  $k_{\rm max} \sim 0.2 \hinvMpc$. With roughly 5 times the $k_{\rm max}$, more bispectrum modes are analyzed in the case of MegaMapper, thus providing significantly more SNR compared to DESI. The one-loop bispectrum requires 3 times more EFT parameters to be marginalized over compared to those that enter the tree-level bispectrum. Hence, there is a competing effect from the increased signal-to-noise ratio to the enlargement of the parameter space when analyzing the one-loop bispectrum. Nevertheless, the analysis of the one-loop bispectrum is expected to provide stronger constraints compared to the tree level analysis. We note that as a function of redshift, higher redshift surveys such as MegaMapper will push the non-linear scale higher, granting more perturbative reach to the same 1-loop model and thereby allowing more modes to be analyzed.  At the same time, the lower number of tracers at higher redshifts also leads to higher shot-noise and thus suppresses the signal to noise ratio.

\begin{figure}[h!]\centering
\hspace*{-1.5em}
\begin{tikzpicture}
\node [above right,inner sep=0] (image) at (0,0) {
\includegraphics[width=\columnwidth]{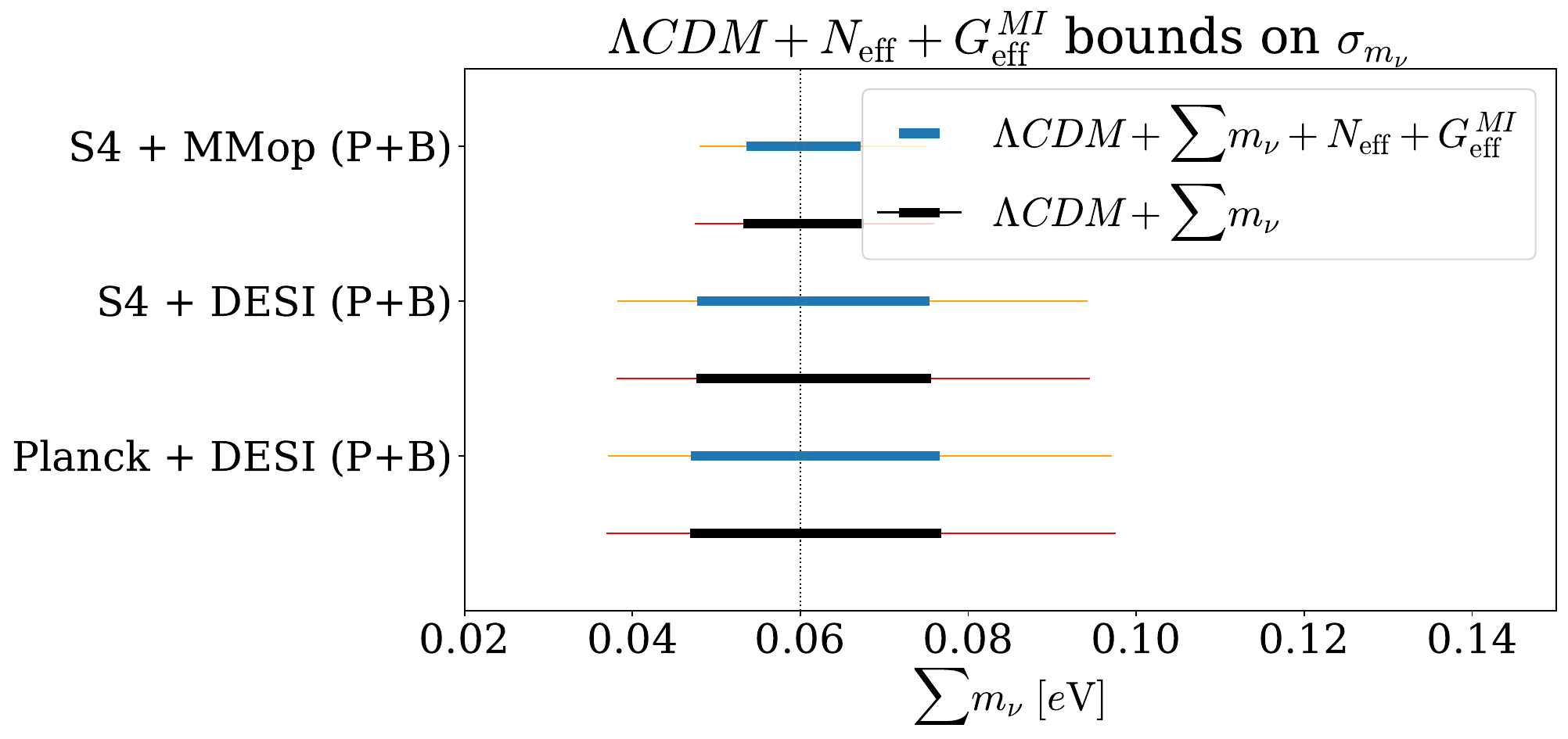}
};
\begin{scope}[
x={($0.1*(image.south east)$)},
y={($0.1*(image.north west)$)}]
\filldraw [fill=white,fill opacity=1, draw=white] (0,3) rectangle (2.95,9);
\node[fill=white,align=center] at (6.5,0.45) {$\Smnu$ [eV]};
\node[fill=White, rounded corners=3mm, fill opacity=1, text opacity=1,align=center] at (6.3,9.65) {\hspace{5em}Bounds on $\Smnu$\hspace{5em}};
\begin{small}
\node[anchor=east] at (3,7.5) {S4+MegaM. (P+B)};
\node[anchor=east] at (3,5.3) {S4+DESI (P+B)};
\node[anchor=east] at (3,3.2) {Pl.+DESI (P+B)};
\draw[Gray!80] (2.95,6.5)--(9.92,6.5);
\draw[Gray!80] (2.95,4.4)--(9.92,4.4);
\end{small}
\end{scope} 
\end{tikzpicture}
\caption{
Neutrino mass constraints of $\LCDM + \Smnu + \Neff +\Geff$ in comparison to $\LCDM + \Smnu$ for S4 + MegaMapper, S4 + DESI and Planck + DESI experiments when analyzing both the galaxy power spectrum and bispectrum at the one-loop order. Here we assume a fiducial mass centered on minimal mass in NO.}
\label{fig:NeffGeff_PB_bounds}
\end{figure}

\paragraph{Robustness against variations of the theoretical model.}
In order to test the robustness of future neutrino mass constraints, we consider several $\LCDM + \Smnu$ extensions with new degrees of freedom that can be partially degenerate with $\Smnu$. 
We check if those can be partially degenerate with $\Smnu$, therefore potentially degrading the constraints when marginalizing over the `new physics' or introducing some level of bias when not accounted for. With two new parameters $\Neff$ and $\Geff$ in the SI$\nu$ model 
we find the bounds do not significantly change when compared to the baseline model as shown in~\cref{fig:NeffGeff_PB_bounds}. This is consistent with the Planck + BOSS analysis shown in ~\cite{He:2023oke}. We show in~\cref{tab:all_bounds} the summary of our forecasts for all $\LCDM+ \Smnu$ extensions we consider.

\begin{table}[h!]\centering%
\begin{tabular}{lccccc}
\toprule
\hfill\multirow{2}{*}{$\sigma_{\Smnu} [\meV]$} & 
\multirow{2}{*}{\small Planck} & \multirow{2}{*}{S4}  & \raisebox{-0.25\height}{Planck+}  & \raisebox{-0.25\height}{\small S4+} & \raisebox{-0.25\height}{\small S4+} \\ \addlinespace[.25em]
 & & & \raisebox{0.25\height}{\small DESI} & \raisebox{0.25\height}{\small DESI} & \raisebox{0.25\height}{\small MegaM.} \\ \midrule
$\LCDM$+$\Smnu$ & $140$ & $40$ & $15$  & $14$ & $7$  \\ 
\tabhspace+$\Neff$ & $150$ & $40$ & $15$ & $14$ & $7$ \\ 
\tabhspace+$\Neff$+$\Geff^\text{MI}$ & $140$ & $42$ &  $16$ & $14$ & $7$ \\ 
\tabhspace+$\delta m_e$ & $130$ & $43$ & $24$ & $15$ & $8$ \\ 
\tabhspace+$\Omega_k$ & $140$ & $56$ & $20$ & $17$  & $7$ \\ 
\bottomrule
\end{tabular}
\caption{\PP{Forecasted uncertainty on neutrino masses for NO scenario $\Smnu = 60\, \meV$, quoted as the average one-sigma bound: $\sigma_{m_\nu} = (\sigma_+ - \sigma_-)/2$.} }
\label{tab:all_bounds}
\end{table}

For most scenarios, the inclusion of the DESI $P+B$ analysis improves neutrino mass constraints by a factor of 2 over Planck alone. For all scenarios, we find that when combining CMB and LSS data, enlarging the parameter space do not deteriorate significantly. For example, with $\SInu$, the Planck and S4 bounds relax by less than 5\% and the Planck + DESI and S4 + MegaMapper bounds remain unchanged. There is however one exception in the case of varying electron mass, where the constraint is relaxed in the case of Planck + DESI due to large degeneracies in the posterior distribution. With S4 + DESI, however, the neutrino mass constraint becomes as robust in the varying electron mass scenario as other extensions. Swapping DESI with MegaMapper provides a further gain of a factor of 2 on the neutrino mass constraints across all $\LCDM$ extensions.

\begin{figure}[h]\centering
\includegraphics[width=.9\columnwidth,trim = 0 3 0 15,clip]{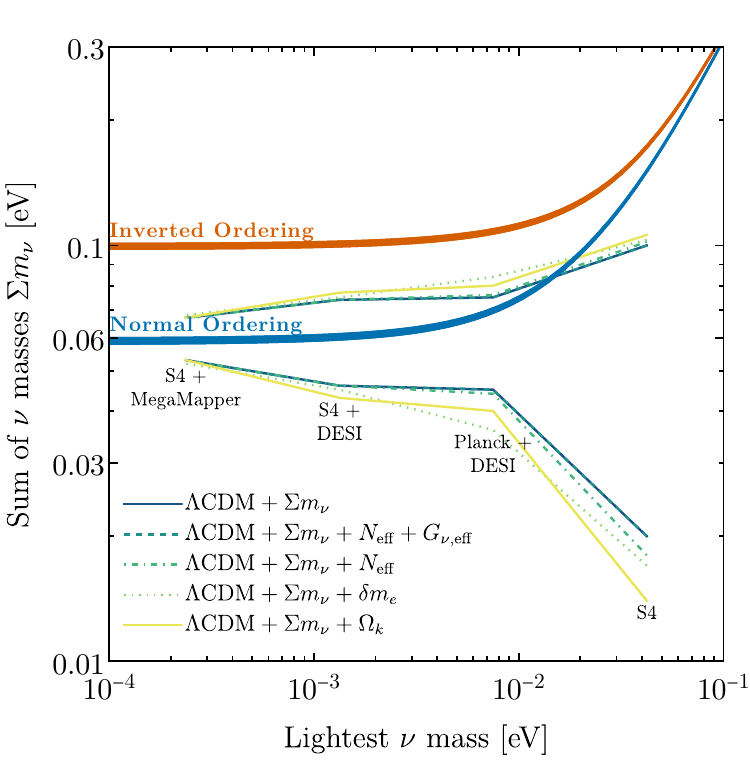}
\caption{Summary of $\Smnu$ constraints assuming minimal mass in NO for various experiments and theory models. For all combinations we show the 68\% CI bound when analyzed with the full one-loop predictions for the power spectrum and bispectrum in the EFTofLSS.}
\label{fig:mnu_All-Models}
\end{figure}

While S4 does not provide significantly more constraining power compared to Planck when combined with LSS for the nominal $\LCDM + \Smnu$ case the joint analysis of S4 + LSS gives robust neutrino mass constraints across all the $\LCDM$ extensions, as can be seen from~\cref{fig:NeffGeff_PB_bounds,fig:mnu_All-Models}. In comparison, when S4 is replaced with Planck, we see that the neutrino constraint for varying electron mass and curvature significantly deteriorates.

We note that similar forecasts for Euclid and CMB were presented in~\cite{Euclid:2024imf}, using however the information from the power spectrum only, with an approximate modeling of nonlinearities. 
In comparison, we also make sure of the information of the bispectrum on scales enabled by the one loop, yielding tighter constraints at \PP{$\sim 30 \%$} when analyzing the combination of DESI + Planck compared to Euclid + Planck at the 1-$\sigma$ level. Yet, the neutrino constraints of Euclid + S4, found to be $\sigma_{\Smnu} = 0.016\, \eV$ in~\cite{Euclid:2024imf}, is similar to DESI + S4 as shown in~\cref{tab:all_bounds}.  

\paragraph{Correlations between $\Smnu$ and new-physics parameters.}

The neutrino mass affects both the CMB and the matter power spectrum by changing the total non-relativistic matter density at late times,
$\delta_{m} = (1-f_\nu)\delta_{cb} + f_{\nu}\delta_{\nu}$, besides the background evolution that affects the growth of structures.
In the baseline $\LCDM + \Smnu$ model, $\Omega_{m}$ and $\omega_{b}$ are varied, and thus the fractional matter density $\Omega_{m}$ is sensitive to changes in the neutrino dynamics. 
Furthermore, in the CMB, both $\Smnu$ and $h$ enter the determination of the angular diameter distance at recombination $d_{A}(z_{\rm rec})$~\cite{ParticleDataGroup:2020ssz}. 
Given that degeneracies between $\Smnu$ and other cosmological parameters arise from different sources in the CMB and the matter power spectrum, we study in particular the correlation of $\Omega_{m}$ and $\Smnu$ when considering CMB alone, LSS alone, and CMB + LSS in combination. 
Looking at just the CMB posteriors, the direction of the degeneracy between $\Omega_{m}$ and $m_{\nu}$ is the same across all models, as shown in \cref{tab:Om_mnu_corr}.

\begin{table}[h!]\centering
\begin{tabular}{lcc@{\quad }cc}
\toprule
\hfill\multirow{2}{*}{$\rho(\Omega_{m}, \Smnu)$} & Planck & \raisebox{-.25\height}{Planck} & S4 & \raisebox{-.25\height}{S4+}\\ \cline{2-2} \cline{4-4}\addlinespace[.25em]
 & DESI  & \raisebox{.25\height}{+\,DESI} & MegaM. &  \raisebox{.25\height}{MegaM.}\\ \midrule
 
\multirow{2}{*}{$\Lambda$CDM+$\Smnu$} & \gr{0.86} & \multirow{2}{*}{\gr{0.3}} & \gr{0.9} & \multirow{2}{*}{\gr{0.6}} \\
 & \gr{0.88}   &\multirow{-2}{*}{\gr{0.3}}  & \gr{0.73} & \multirow{-2}{*}{\gr{0.6}} \\ \midrule
 
\multirow{2}{*}{\tabhspace+$\Neff$} & \gr{0.83}  & \multirow{2}{*}{\gr{0.18}} & \gr{0.8} & \multirow{2}{*}{\gr{0.5}} \\
   & \gr{0.87}  &  \multirow{-2}{*}{\gr{0.18}} & \gr{0.86} &  \multirow{-2}{*}{\gr{0.5}} \\ \midrule
   
\multirow{2}{*}{\tabhspace+$\Neff$+$\Geff$} & \gr{0.76}  & \multirow{2}{*}{\gr{0.05}}  & \gr{0.8} & \multirow{2}{*}{\gr{0.37}}  \\ 
  & \gr{0.88} &  \multirow{-2}{*}{\gr{0.05}} & \gr{0.56}  & \multirow{-2}{*}{\gr{0.37}}  \\ \midrule
  
\multirow{2}{*}{\tabhspace+$\delta_{m_{e}}$} & \gr{0.17}  & \multirow{2}{*}{$-$\gr{0.26}} & \gr{0.15} & \multirow{2}{*}{\gr{0.09}} \\
 & \gr{0.69} &  \multirow{-2}{*}{$-$\gr{0.26}}  & \gr{0.57}  & \multirow{-2}{*}{\gr{0.09}} \\ \midrule
 
\multirow{2}{*}{\tabhspace+$\Ok$} & \gr{0.69}  & \multirow{2}{*}{\gr{0.48}}  & \gr{0.8} & \multirow{2}{*}{\gr{0.64}}  \\
 & \gr{0.88} &  \multirow{-2}{*}{\gr{0.48}}  & \gr{0.93} & \multirow{-2}{*}{\gr{0.64}} \\ \bottomrule
\end{tabular}
\caption{\PP{Correlation coefficients $\rho(\Omega_{m}, \Smnu)$ for different models and experiments.
The columns show the values forecast for a CMB (Planck or S4) and LSS (DESI or MegaM.) surveys separately, and in combination.}\\
}
\label{tab:Om_mnu_corr}
\end{table}

\begin{figure}[h]\centering
\hspace*{-1.5em}
\begin{tikzpicture}
\node [above right,inner sep=0] (image) at (0,0) {
\includegraphics[width=1.05\columnwidth,trim = 5 3 0 5,clip]{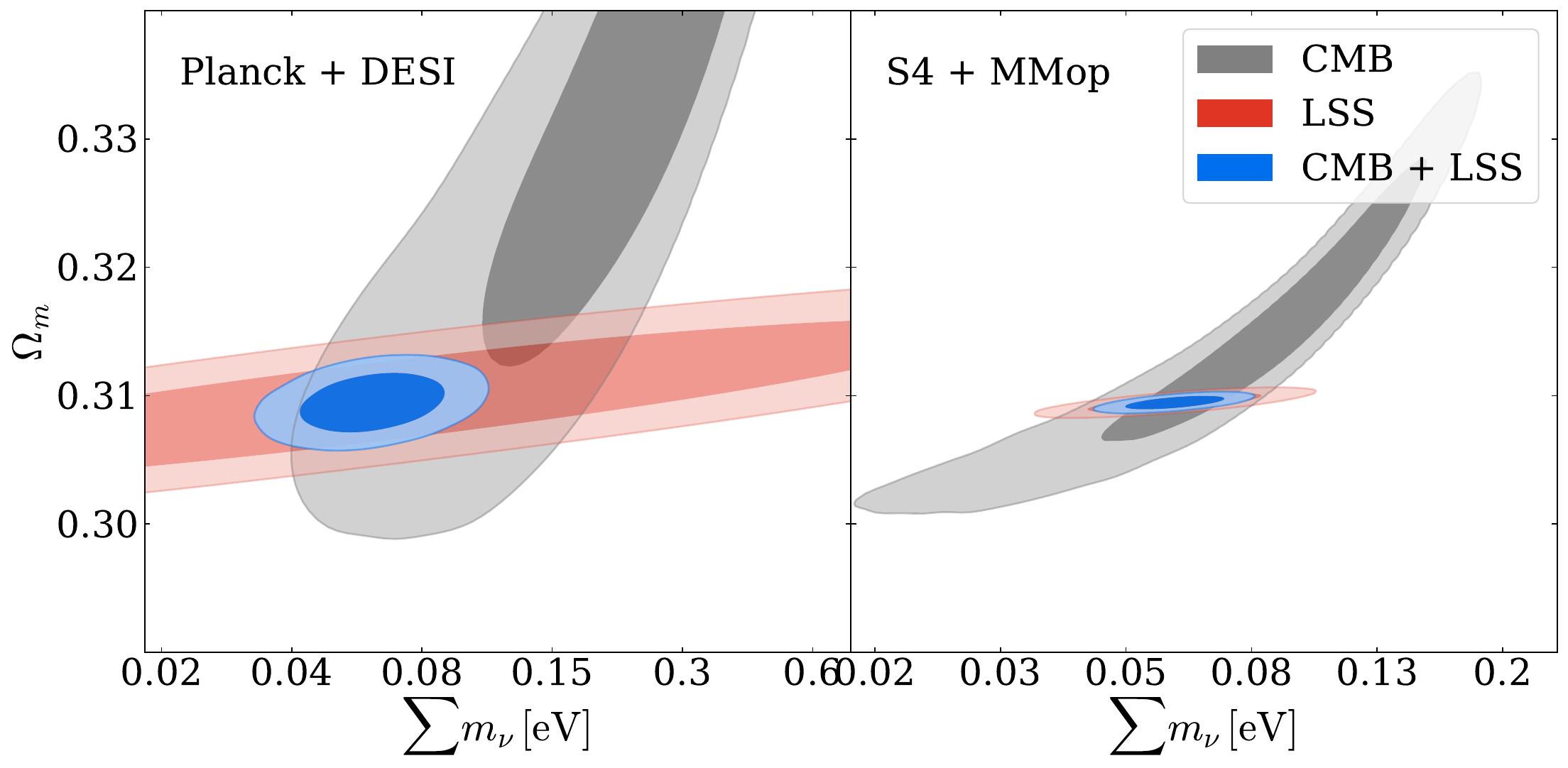}
};
\begin{scope}[x={($0.1*(image.south east)$)},
y={($0.1*(image.north west)$)}]
\node[fill=white,align=center] at (6.4,9.1) {\scriptsize S4+MegaM.};
\end{scope} 
\end{tikzpicture}
\caption{Posteriors for $\Omega_{m}$ vs $\Smnu$ in the base $\LCDM + \Smnu$ model. Note that the posterior distribution is not ellipsoidal due to the projection of neutrino mass from log to linear space. \textit{Left}: Posteriors for Planck only, DESI only, and Planck + DESI analysis. \textit{Right}: Posteriors for S4, MegaMapper only, and S4 + MegaMapper analysis. Notice the different ranges of $\Smnu$ in the two plots.
}
\label{fig:2dOm_mnu}
\end{figure}

The degeneracy directions of $\Omega_{m}$ and $\Smnu$ differ more significantly between Planck and DESI compared to S4 and MegaMapper in~\cref{fig:2dOm_mnu}. 
When combining CMB and LSS, this degeneracy is then reduced more significantly for Planck + DESI than for S4 + MegaMapper.
The non-Gaussian nature of the CMB posterior due to the projection from $\Smnu$ to $\log(\Smnu)$ leads to a degeneracy-breaking pattern that is less intuitive than if the posteriors were Gaussian. 
For example, in the case of the model $\LCDM+\Smnu +\delta m_{e}$, the Planck + DESI combination yields a negative correlation due to the non-Gaussian shape of the Planck posterior, despite individually having positive correlations. Nevertheless, the combined analysis of CMB + LSS does alleviate some of the parameter degeneracies with $\Smnu$, resulting in more robust neutrino bounds.

\begin{figure}[h]\centering
\hspace*{-1.5em}
\begin{tikzpicture}
\node [above right,inner sep=0] (image) at (0,0) {
\includegraphics[width=1.05\columnwidth,trim = 5 3 0 5,clip]{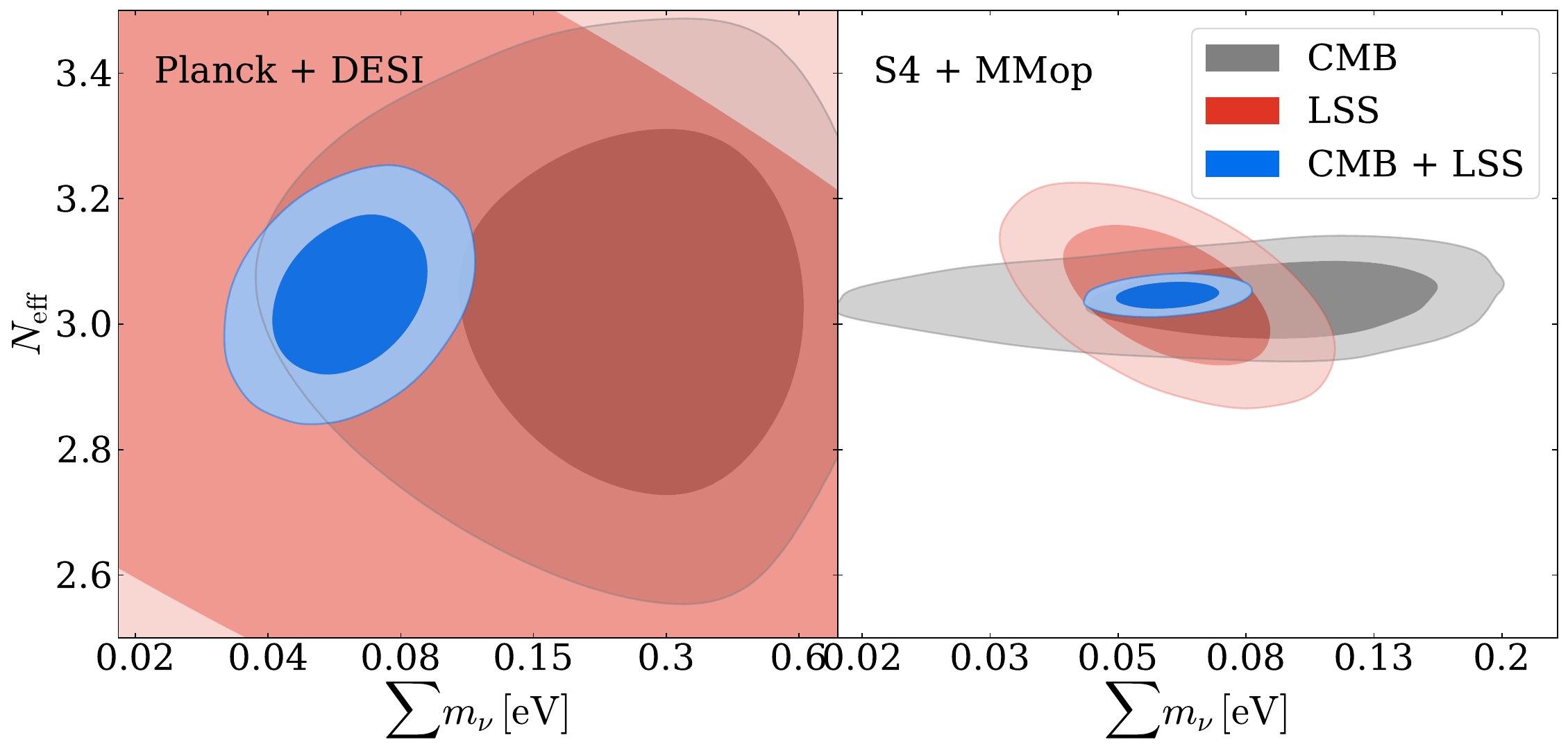}};
\begin{scope}[x={($0.1*(image.south east)$)},
y={($0.1*(image.north west)$)}]
\node[fill=white,align=center] at (6.4,9.1) {\scriptsize S4+MegaM.};
\end{scope} 
\end{tikzpicture}
\caption{Same as \cref{fig:2dOm_mnu}, for $\Neff$ vs $\Smnu$ in the $\LCDM + \Smnu+\Neff$ model.
}
\label{fig:2dNur_mnu}
\end{figure}
\begin{figure}[h]\centering
\hspace*{-1.5em}
\begin{tikzpicture}
\node [above right,inner sep=0] (image) at (0,0) {
\includegraphics[width=1.05\columnwidth,trim = 5 3 0 5,clip]{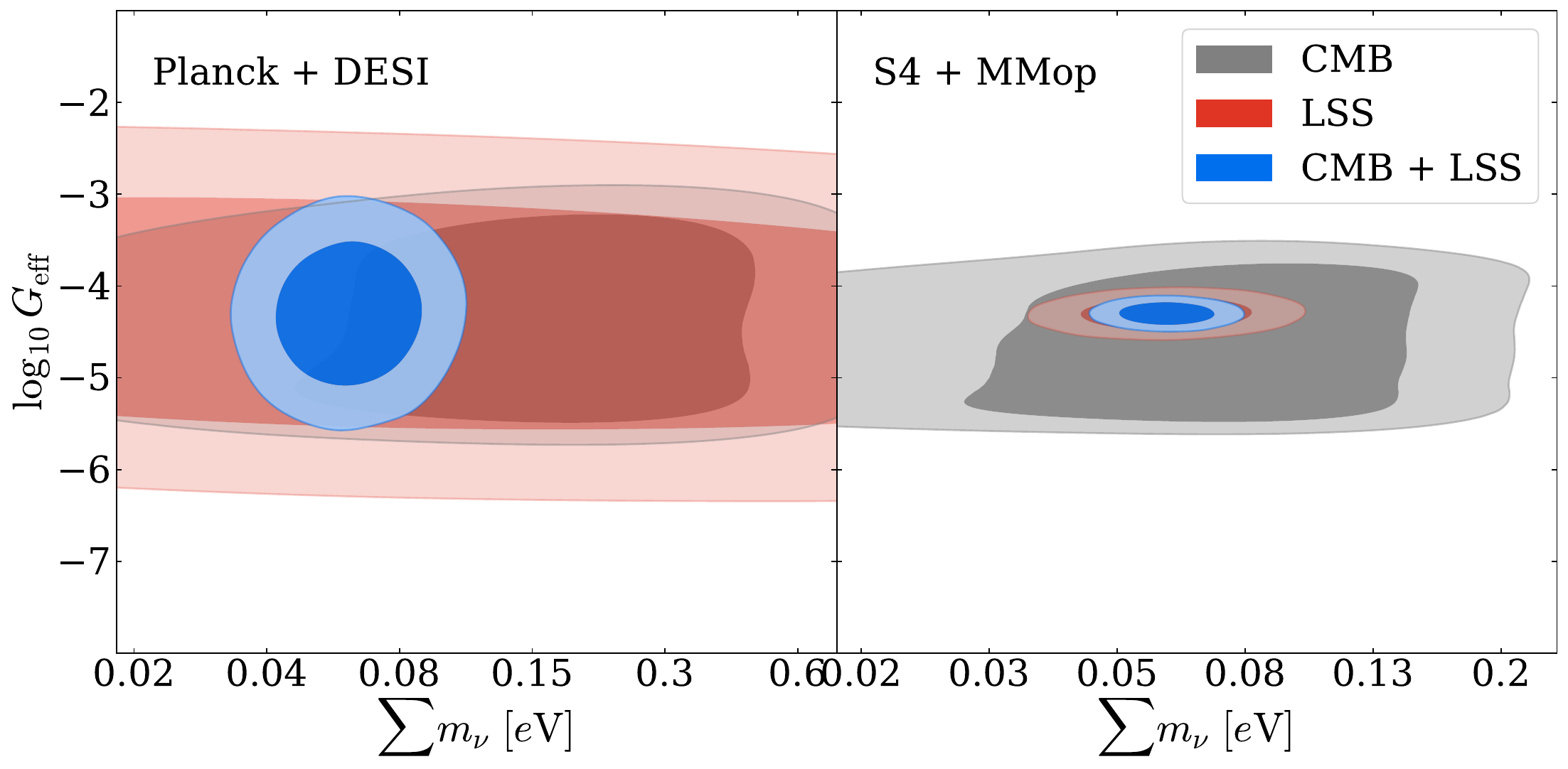}
};
\begin{scope}[x={($0.1*(image.south east)$)},
y={($0.1*(image.north west)$)}]
\node[fill=white,align=center] at (6.4,9.1) {\scriptsize S4+MegaM.};
\end{scope} 
\end{tikzpicture}
\caption{Same as \cref{fig:2dOm_mnu}, for $\log_{10}(\Geff/\MeV^{-2})$ vs $\Smnu$ in the $\LCDM + \Smnu+\Neff + \Geff$ model.}
\label{fig:2dGeff_mnu}
\end{figure}
\begin{figure}[h]\centering
\hspace*{-1.5em}
\begin{tikzpicture}
\node [above right,inner sep=0] (image) at (0,0) {
\includegraphics[width=1.05\columnwidth,trim = 5 3 0 5,clip]{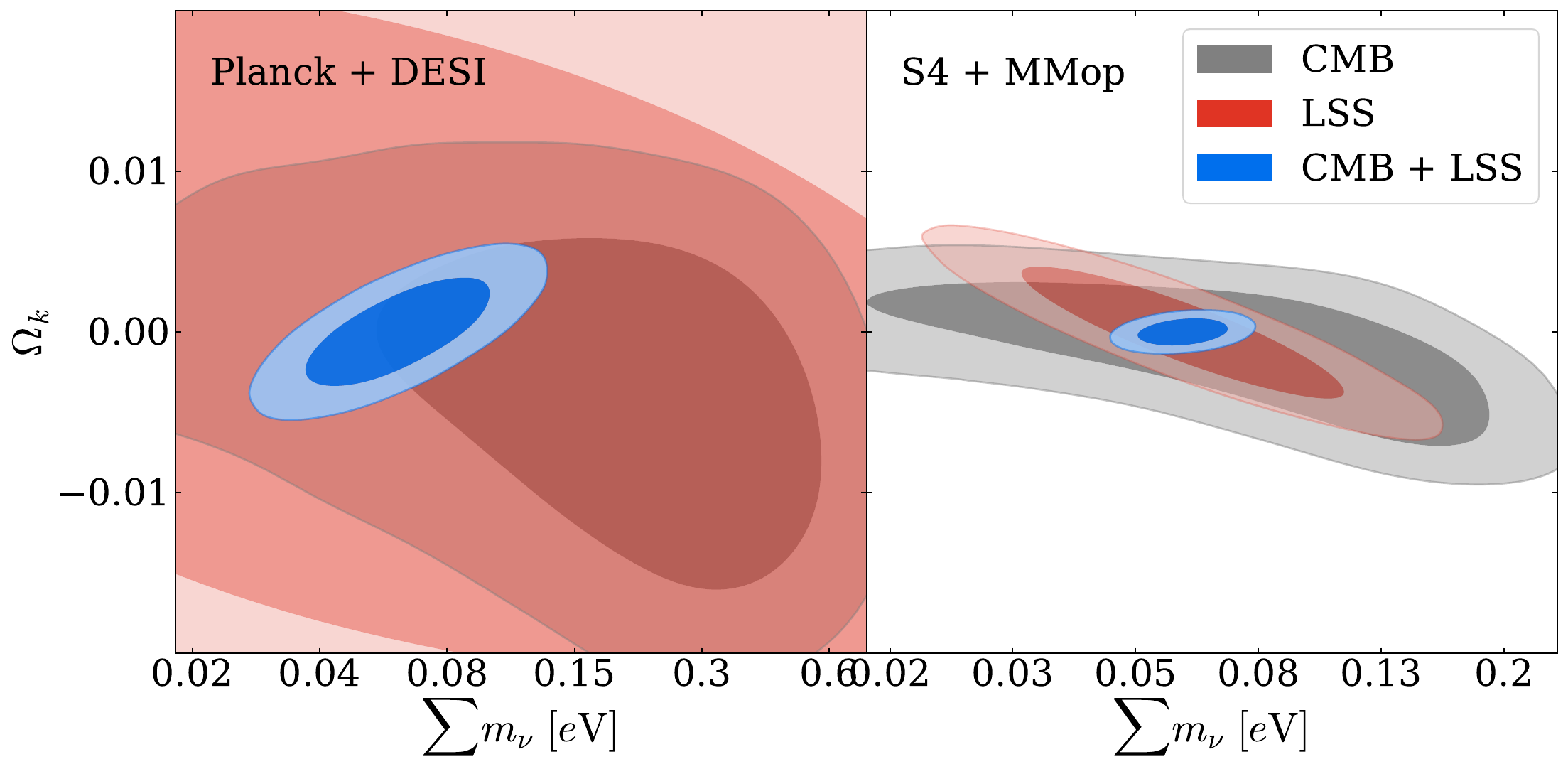}
};
\begin{scope}[x={($0.1*(image.south east)$)},
y={($0.1*(image.north west)$)}]
\node[fill=white,align=center] at (6.5,9.1) {\scriptsize S4+MegaM.};
\end{scope} 
\end{tikzpicture}
\caption{Same as \cref{fig:2dOm_mnu}, for $\Ok$ vs $\Smnu$ in the $\LCDM +\Smnu+ \Ok$ model.
}
\label{fig:2dOmega_k_mnu}
\end{figure}
\begin{figure}[h]\centering
\hspace*{-1.5em}
\begin{tikzpicture}
\node [above right,inner sep=0] (image) at (0,0) {
\includegraphics[width=1.05\columnwidth,trim = 5 3 0 5,clip]{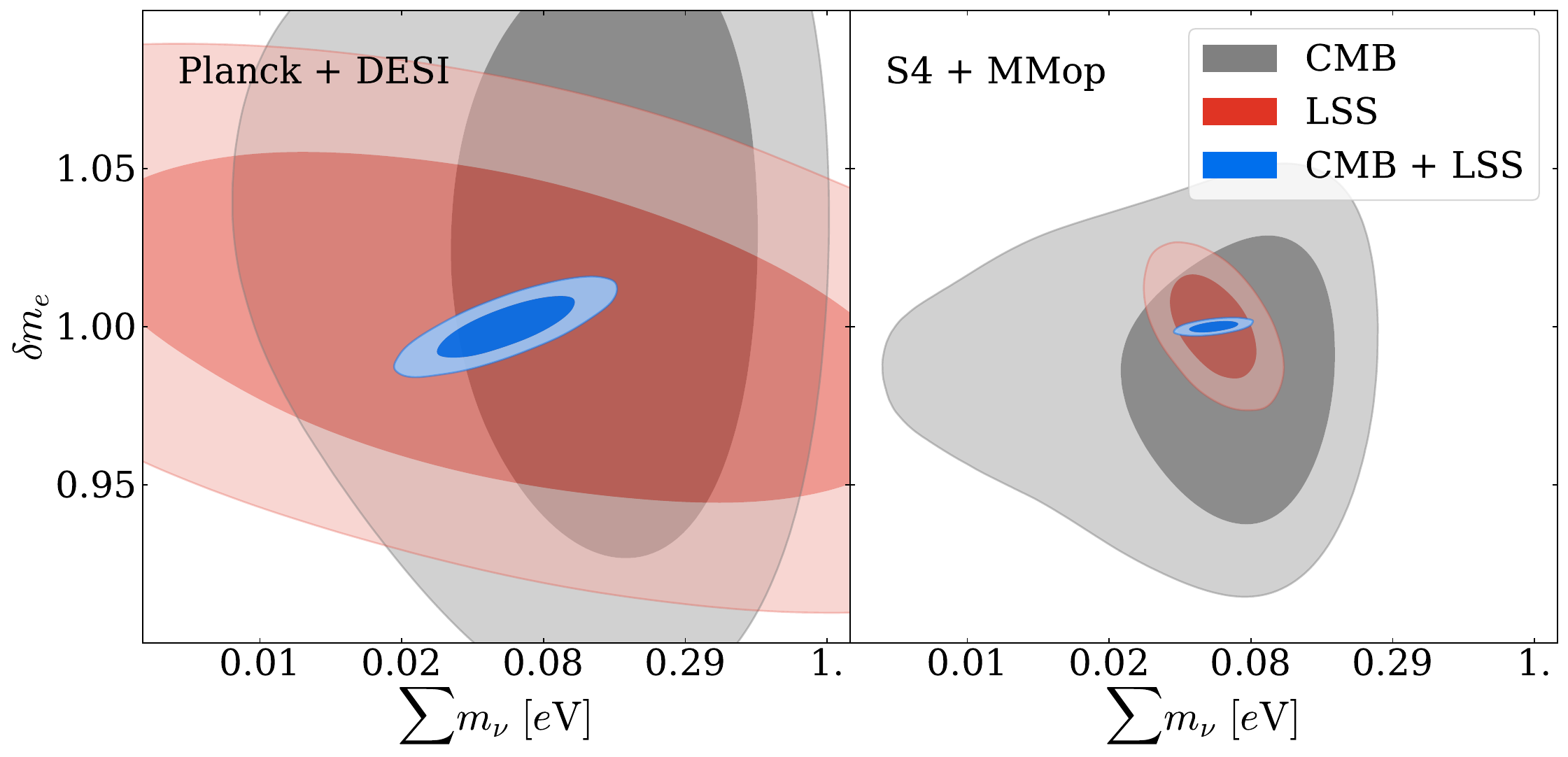}
};
\begin{scope}[x={($0.1*(image.south east)$)},
y={($0.1*(image.north west)$)}]
\node[fill=white,align=center] at (6.4,9.1) {\scriptsize S4+MegaM.};
\end{scope} 
\end{tikzpicture}
\caption{Same as \cref{fig:2dOm_mnu}, for $\delta m_e$ vs $\Smnu$ in the $\LCDM +\Smnu+ \delta m_e$ model. Notice the wider range in $\Smnu$ with respect to the previous plots.}
\label{fig:2dme_mnu}
\end{figure}

The Fisher study thus far has been evaluated on a fiducial cosmology that assumes no new physics, i.e.\;$\Geff = 0, \, \Neff = 0, \, \Ok = 0,\, \delta m_{e} = 1$. Without having to evaluate the Fisher matrices with a new fiducial, one can estimate the shift in the expected neutrino mass given a shift in the new physics parameter by examining their correlations
\footnote{We note that this is only an estimate of the parameter shifts using the 2D posterior and correlation and a full MCMC analysis with a shifted fiducial is required for a more comprehensive and quantitative study of the parameter shift.}
as shown in ~\cref{fig:2dGeff_mnu,fig:2dNur_mnu,fig:2dme_mnu,fig:2dOmega_k_mnu}, and summarised in \cref{tab:NP_mnu_corr}. 

\begin{table}[h!]\centering
\begin{tabular}{ccc@{\quad }cc}
\toprule
\hfill\multirow{2}{*}{$\rho(\cdot, \Smnu)$} & Planck & \raisebox{-.25\height}{Planck} & S4 & \raisebox{-.25\height}{S4+}\\ \cline{2-2} \cline{4-4}\addlinespace[.25em]
 & DESI  & \raisebox{.25\height}{+\,DESI} & MegaM. &  \raisebox{.25\height}{MegaM.}\\ \midrule

\multirow{2}{*}{$\Neff$} & {$-$\gr{0.05}}  & \multirow{2}{*}{\gr{0.3}} & \gr{0.14} & \multirow{2}{*}{\gr{0.21}} \\
   & {$-$\gr{0.73}}  &  \multirow{-2}{*}{\gr{0.3}} & {$-$\gr{0.5}} &  \multirow{-2}{*}{\gr{0.21}} \\ \midrule
   
\multirow{2}{*}{$\Geff$} & \gr{0.08}  & \multirow{2}{*}{\gr{0.03}}  & \gr{0.11} & \multirow{2}{*}{$-$\gr{0.02}}  \\ 
  & {$-$\gr{0.23}} &  \multirow{-2}{*}{\gr{0.03}} & \gr{0.08}  & \multirow{-2}{*}{$-$\gr{0.02}}  \\ \midrule
  
\multirow{2}{*}{$\delta{m_{e}}$} & {$-$\gr{0.02}}  & \multirow{2}{*}{\gr{0.71}} & \gr{0.02} & \multirow{2}{*}{\gr{0.48}} \\
 & {$-$\gr{0.49}} &  \multirow{-2}{*}{\gr{0.71}}  & {$-$\gr{0.49}}  & \multirow{-2}{*}{\gr{0.48}} \\ \midrule
 
\multirow{2}{*}{$\Ok$} & {$-$\gr{0.36}}  & \multirow{2}{*}{\gr{0.71}}  & {$-$\gr{0.6}} & \multirow{2}{*}{\gr{0.26}}  \\
 & {$-$\gr{0.44}} &  \multirow{-2}{*}{\gr{0.71}}  & {$-$\gr{0.89}} & \multirow{-2}{*}{\gr{0.26}} \\ \bottomrule
\end{tabular}
\caption{\PP{Correlation coefficients of $\Smnu$ with other new physics parameters $\rho(\cdot, \Smnu)$ for different models and experiments.
The columns show the values forecast for a CMB (Planck or S4) and LSS (DESI or MegaM.) surveys separately, and in combination.}}
\label{tab:NP_mnu_corr}
\end{table}

For $\LCDM$ + $\Neff$ and $\Geff$, there is little correlation between the new physics parameter and $\Smnu$ when analyzing CMB + LSS. As such, the central neutrino mass is not expected to shift from the fiducial even when the new physics parameter is shifted by 1$\sigma$. 
In the case of $\Omega_{k}$ and $\delta m_{e}$, there is a correlation \PP{$\sim 0.7$} with $\Smnu$ for Planck + DESI. This implies a shift in the central value of $\Smnu$ of about \PP{$0.7\sigma$} along the direction of correlation if the measured central value of the new physics parameter deviates from its fiducial by $1\sigma$. 
This correlation is strongly reduced to \PP{$\sim 0.5$} when analyzing S4 + MegaMapper.
One can notice though, from the comparison of \cref{tab:NP_mnu_corr} and \cref{fig:2dGeff_mnu,fig:2dNur_mnu,fig:2dme_mnu,fig:2dOmega_k_mnu}, that in some cases the correlation parameter can be very small not because of a small covariance between the two parameters (the numerator), but because of a large standard deviation of either of the parameters (the denominator).

\begin{table}[h!]\centering
\begin{tabular}{ccc@{\quad }cc}
\toprule
$\Delta\Smnu$ & Planck & \raisebox{-.25\height}{Planck} & S4 & \raisebox{-.25\height}{S4+}\\ \cline{2-2} \cline{4-4}\addlinespace[.25em]
[meV] & DESI  & \raisebox{.25\height}{+\,DESI} & MegaM. &  \raisebox{.25\height}{MegaM.}\\ \midrule

\multirow{2}{*}{$\Neff$} & {$-${2}}  & \multirow{2}{*}{\gs{4.8}} & {4.2} & \multirow{2}{*}{\gs{1.5}} \\
   & {$-${470}}  &  \multirow{-2}{*}{\gs{4.8}} & {$-${7.5}} &  \multirow{-2}{*}{\gs{1.5}} \\ \midrule
   
\multirow{2}{*}{$\Geff$} & {6.4}  & \multirow{2}{*}{\gs{0.6}}  & {6.6} & \multirow{2}{*}{$-$\gs{0.1}}  \\ 
  & {$-${150}} &  \multirow{-2}{*}{\gs{0.6}} & {1}  & \multirow{-2}{*}{$-$\gs{0.1}}  \\ \midrule
  
\multirow{2}{*}{$\delta{m_{e}}$} & {$-${1.6}}  & \multirow{2}{*}{\gs{19}} & {1.2} & \multirow{2}{*}{\gs{4.1}} \\
 & {$-${160}} &  \multirow{-2}{*}{\gs{19}}  & {$-${7.2}}  & \multirow{-2}{*}{\gs{4.1}} \\ \midrule
 
\multirow{2}{*}{$\Ok$} & {$-${25}}  & \multirow{2}{*}{\gs{14}}  & {$-${42}} & \multirow{2}{*}{\gs{1.7}}  \\
 & {$-${130}} &  \multirow{-2}{*}{\gs{14}}  & {$-${22}} & \multirow{-2}{*}{\gs{1.7}} \\ \bottomrule
\end{tabular}
\caption{
\PP{Forecasted 1-$\sigma$ shifts of $\Smnu$ (in meV) from correlation with new physics parameters, defined as $\Delta \Smnu\equiv \text{corr}(\cdot, \Smnu)\sigma_{\Smnu}$ for different models and experiments. 
The columns show the values forecast for a CMB (Planck or S4) and LSS (DESI or MegaM.) surveys separately, and in combination.
For comparison, these shifts can be compared with the respective sensitivities on $\sigma_{\Smnu}$ in \cref{tab:all_bounds}.}
}
\label{tab:NP_mnu_shift}
\end{table}

For this reason, in \cref{tab:NP_mnu_shift} we show the quantity that we are most interested in, that is the approximate directional shift in the measured $\Smnu$ when the observed value of a new physics parameter shifts by 1-$\sigma$. The shift $\Delta \Smnu (X)$ is calculated from its correlation with a new physics parameter $X$ as $\Delta \Smnu (X) = \rho(X,\Smnu)\sigma_{\Smnu}$. LSS experiments alone can have large shifts in $\Smnu$ compared to Planck or S4. In particular, the $\Neff$ and $\Geff$ extensions have wider shifts due to the large degeneracy between $\Neff$ and $\Smnu$. However, the joint CMB and LSS analysis is much more robust to the 1$\sigma$ shifts in new physics parameters as we see the shift in $\Smnu$ is at most less than the 1$\sigma$ bounds on $\Smnu$ reported in \cref{tab:all_bounds}.


\section{Discussions and Conclusions}
\label{sec:discussion}
In this paper, we employ the state-of-the-art EFTofLSS, including both the power spectrum and the bispectrum at one loop in perturbation theory, to forecast the sensitivity of ongoing and future galaxy surveys (DESI, MegaMapper) to $\Smnu$ and its parameter degeneracies, with a particular focus on the impact of new physics (both $\LCDM$ and BSM extensions).
We include neutrino self-interactions, variations of SM parameters as $m_e$, an extra contribution $\DNeff$ of decoupled relativistic species, and a curvature component $\Ok$.
These models identify plausible directions for new physics, and focus on dynamics which are partially degenerate with the cosmological effects of massive neutrinos.
Our purpose is to contribute to the crucial assessment of the incoming cosmological measurement of $\Smnu$, and of its robustness to the effects of new physics in particle physics or cosmology.
We can summarise our results along two main directions.

\paragraph{Forecast of the sensitivity $\sigma_{\Smnu}$.}
We quantify the sensitivity $\sigma_{\Smnu}$ for various theoretical models and future surveys, after marginalising over all the other parameters (see \cref{fig:mnu_bands,tab:all_bounds}).
The combination of Planck with the ground-based telescope DESI is expected to reach within the next $\sim5$ years \PP{$\sigma_{\Smnu}=15 \meV$}, enough to claim a non-zero $\Smnu$ at almost $4\sigma$ and a discrimination between NO and IO above $2\sigma$ (in case of NO and minimal $\Smnu$). 
This result will be excitingly close to the timescale expected for the determination of the mass ordering from oscillation experiments \cite{Cabrera:2020ksc}, and should precede by a few years the achievement of a comparable sensitivity with the Euclid satellite mission.

To get an impression of the improvement determined by the LSS analysis, the recent analysis \cite{Shao:2024mag} of current constraints from Planck+ACT+SNe with DESI BAO gives a sensitivity $\sigma_{\Smnu}\approx 70 \meV$, four times larger than what we forecast with the $P+B$ one-loop analysis of DESI 5-years in combination with Planck.
Another point of comparison are the recent results from DESI 1-year using the one-loop power spectrum~\cite{DESI:2024hhd}, that lead to a $2\sigma$ constraint $\Smnu <  71 \meV$ in combination with CMB. 
The future sensitivity for S4+MegaMapper reaches \PP{$\sigma_{\Smnu}=7\meV$}, which implies a $5\sigma$ sensitivity on the ordering (for the minimal mass in NO). The bispectrum plays a significant role in reaching this threshold sensitivity, as it strengthens by \PP{33\%} the result from the $P_\text{1-loop}$ analysis of MegaMapper (and \PP{$66\%$} of this gain comes from the reach of the $B_\text{1-loop}$ compared to $B_\mathrm{tree}$).

\paragraph{Comparison with previous forecasts for Euclid and DESI.}
We summarise in \cref{tab:comparison literature} the forecasts in the literature about the measurement of $\Smnu$ in DESI and Euclid for the baseline model and with $\Neff$.
\begin{table}[h!]\centering
\begin{tabular}{c>{\small}l>{\small}l@{\ }c|c}
\toprule
& & & \multicolumn{2}{c}{$\sigma_{\Smnu}$ [meV]}\\ \midrule
\multirow{2}{*}{Ref.} & \multirow{2}{*}{\normalsize Forecast} & \multirow{2}{*}{\normalsize Model} & {\scriptsize Pl.+DESI} & {\scriptsize Pl.+Euclid} \\ \cline{4-5}
& & & {\scriptsize S4+DESI} & {\scriptsize S4+Euclid} \\ \midrule
\multirow{4}{*}{\cite{Brinckmann:2018owf}}
& \multirow{4}{8.2em}{DESI BAO / Euclid $P_{\rm Kaiser}$+NL marg., mock MCMC} 
& \multirow{2}{*}{$\Lambda$CDM+$\Smnu$} & 44 & 20 \\ \cline{4-5}
& & & 19 & 12 \\ \addlinespace[.25em]
& & \multirow{2}{*}{\tabhspace+$\Neff$} & 47 & 23 \\ \cline{4-5}
& & & 21 & 14 \\ \midrule
\multirow{2}{*}{\cite{Chudaykin:2019ock}}
& \multirow{2}{8.9em}{$P_\text{1-loop}$, EFTofLSS + AP, mock MCMC} 
& \multirow{2}{*}{$\Lambda$CDM+$\Smnu$} &  & 17 \\ \cline{4-5}
& & & &  \\ \midrule
%
\multirow{4}{*}{\cite{Euclid:2024imf}}
& \multirow{4}{8em}{$P_{\rm Kaiser}$ + FoG + AP, mock MCMC} 
& \multirow{2}{*}{$\Lambda$CDM+$\Smnu$} &  & 23 \\ \cline{4-5}
& & &  & 16 \\ \addlinespace[.25em]
& & \multirow{2}{*}{\tabhspace+$\Neff$} & & 25 \\ \cline{4-5}
& & &  & 16 \\ \midrule
\multirow{4}{2.3em}{this work}
& \multirow{4}{8.3em}{$P_\text{1-loop}$, EFTofLSS (no AP), Fisher} 
& \multirow{2}{*}{$\Lambda$CDM+$\Smnu$} & 17 & \\ \cline{4-5}
& & & 16& \\ \addlinespace[.25em]
& & \multirow{2}{*}{\tabhspace+$\Neff$} & 17 &  \\ \cline{4-5}
& & & 16 &  \\ 
%
\bottomrule
\end{tabular}
\caption{\PP{Comparison between forecasts in the literature for $\sigma_{\Smnu}$ (in meV) for the LSS surveys Euclid and DESI (using the power spectrum only) in combination with CMB surveys (Planck and S4). The main features of the different methodologies are listed in the second column.}
}
\label{tab:comparison literature}
\end{table}

At face value, it seems that the sensitivity of Euclid and DESI in combination with CMB will be very similar, which is expected given that both experiments observe a similar data volume. There are however some differences among these studies.
Clearly, there are caveats in these comparisons: experiment specifications might vary from one study to the other; DESI and Euclid do not target the same tracers, redshift ranges, or sky fraction; and the fiducial parameters are also different.
We can nevertheless highlight aspects related to variations in the modelling and their potential implications for the sensitivity on $\Smnu$. 
Ref.~\cite{Brinckmann:2018owf} uses a Kaiser model while taking into account non-linearities by inflating the covariance with an estimate of the scaling of the theoretical error for Euclid. 
Their forecasts agree with ours at the order-of-magnitude level. 
For DESI, they instead only use BAO information, leading to a constraint almost $3$ times weaker than our forecast. This suggests that the full shape is comparatively very informative for $\Smnu$.
Ref.~\cite{Euclid:2024imf} offers an updated forecast for Euclid, and improves the modelling with a  Finger-of-God (FoG) damping term, smearing of BAO by nonlinearities, AP effects, and corrections to redshift systematic errors.
Taking a conservative $k_{\rm max} \sim 0.25 \hinvMpc$ to mitigate incompleteness in their modelling of nonlinearities, their forecasts further include additional information from Euclid weak lensing and cluster counts. 
Yet, their forecast sensitivities on $\Smnu$ fall close to ours, especially given that our Fisher has a potential $\sim 50\%$ error due to the missing AP effects and the approximate modelling of the covariance.
A potential source of difference between our results and those of Refs.~\cite{Brinckmann:2018owf,Euclid:2024imf} is the treatment of nonlinearities: as discussed in \cref{sec:analytic_fish}, those drastically reduce the correlation between $b_1$ and $\Smnu$, at the cost of additional nuisance parameters to marginalise over.  

Ref.~\cite{Chudaykin:2019ock} uses a similar EFTofLSS modelling as ours, but further includes the AP effect. 
Our final results end up coinciding on $\sigma_{\Smnu}=17\meV$. 
Surprisingly, their addition of $B_\text{tree}$ brings the sensitivity to $13$ meV, while we find that the addition of the $B_\text{1-loop}$ only improves to 
\PP{$15$} meV. 
One difference is that our $k_{\rm max}$ is determined based on extrapolation from actual galaxy data, whereas they use an arbitrarily large $k_{\rm max}$ and add to the covariance an estimate of the one-loop error. 
As a main take-away from this comparison, we observe that the modelling choice is likely to affect the cosmological constraints on $\Smnu$.

\paragraph{Robustness of $\sigma_{\Smnu}$ against new physics.}
The second important aspect of our results concerns the robustness achieved on $\sigma_{\Smnu}$ with respect to new physics, when combining CMB and LSS at such remarkable precision.
The sensitivity $\sigma_{\Smnu}$ on neutrino masses (see \cref{fig:mnu_All-Models,tab:all_bounds}) is pretty robust (with variations up to 50\% across the models we consider, in substantial agreement with e.g.~\cite{Shao:2024mag}) already with Planck+DESI, and will be practically independent of the model for S4+MegaMapper, that achieve enough precision to break parameter degeneracies.
To quantify more precisely this model dependence, within our Fisher forecast we can compute the correlations of $\Smnu$ with the new-physics parameters $X_i$ for each model, and the expected shift $\Delta\Smnu \equiv\text{corr}(\Smnu,X_i)\sigma_{\Smnu}$ in $\Smnu$ due to a $1\sigma$ shift in the new-physics variable $X_i$.
We find that the combination S4+MegaMapper brings the shift $\Delta\Smnu$ down to \PP{$\sim 1- 4\meV$} among all the models we consider. 
Although any selection of theoretical models will be generically incomplete (the recent results from DESI-1y have motivated the reconsideration of physics affecting the inference of $\Smnu$, such as e.g.~decaying neutrinos, long-range dark forces and $f(R)$ models~\cite{FrancoAbellan:2021hdb,Craig:2024tky,Bottaro:2024pcb,Baldi:2013iza,Hu:2014sea}), we believe that our choice was representative of many new parameters that can display degeneracy with $\Smnu$.
The smallness of the shifts $\Delta\Smnu$ for S4+MegaM.\; means that the variance of the measured value of $\Smnu$ should not be significantly impacted by new physics.

\paragraph{Future prospects.}
In conclusion, our study quantifies how ongoing and future LSS surveys (DESI, MegaMapper), when exploited in their full power with the EFTofLSS, allow a remarkably precise determination of the sum of neutrino masses in combination with CMB surveys.
This synergy could allow not only a non-zero detection of $\Smnu$, but also a preliminary determination of the mass ordering in a time span of $\sim5$ years with Planck+DESI, roughly at the same time when the first indications about the mass ordering could come from oscillation measurements (T2K, NOvA, JUNO).
The prospect with S4+MegaMapper is an accurate measurement of $\Smnu$ with a sensitivity as good as \PP{$7-8$} meV, robustly with respect to a wide set of theoretical assumptions on particle physics and cosmology.
Looking forward, a measurement of the first BSM parameter from cosmology will necessarily stand as a precision benchmark for future searches of exciting new physics with the CMB and LSS.

\begin{acknowledgments}
We thank Maria Archidiacono, Emanuele Castorina, Julien Lesgourgues, Massimo Pietroni, Vivian Poulin and Leonardo Senatore for helpful discussions and comments on the draft. 
D.R.\;thanks the participants to the workshop \href{https://indico.cern.ch/event/1375290/overview}{``New physics from Galaxy Clustering III''} (Parma, 4-8 Nov 2024), and the \textsl{Theoretical Cosmology} group at ETH, where this work was presented, for their stimulating feedback. 
We acknowledge use of the publicly available codes {\tt matplotlib}~\cite{Hunter:2007}.
D.R.~was supported at Stanford U.\,by NSF Grant PHY-2014215, DOE HEP QuantISED award \#100495, and the Gordon and Betty Moore Foundation Grant GBMF7946; at U.~of Zurich by the UZH Postdoc Grant 2023 Nr.\,FK-23-130. 
\end{acknowledgments}


\appendix
\section*{Appendices}

\section{Simple analytic Fisher estimates}
\label{sec:analytic_fish}

In this appendix, we provide rough estimates for the $\sigma$-sensitivity on $\Smnu$ from LSS when jointly analysed with CMB. 
These are based on simple analytic expressions following our counting of leading-$f_\nu$ factors in~\cref{sec:EFT-of-LSS}. 
Optimal, in the sense that we rely on a bunch of simplified assumptions to derive them, the bounds presented here allow us to cross-check our results from the full Fisher pipeline presented in the main text, that we find align well. 

We work with the following simplified assumptions. 
First, assuming that all other cosmological parameters are well determined by the combination with CMB, the latter in particular almost fixing $A_s, n_s, \omega_b$, and $\omega_{c,0}$, limits on $\Smnu$ from galaxy clustering data come essentially from the almost flat amplitude suppression on the power spectrum with respect to the no-neutrino case.~\footnote{In reality, it is well-known that the neutrino mass has a strong degeneracy with $\tau$ in the CMB, that we neglect for the purpose of the discussion here. } 
We further assume that the scale-independent growth rate $f_0 \sim \Omega_m^{0.55}$ is fixed by the information from the BAO + CMB.
Also, we leave aside the bispectrum and nonlinear corrections in the power spectrum for now. 
We thus consider a simple Kaiser model for the power spectrum, 
\begin{equation}
P_g(k,\mu) \approx \big(b + (1-3f_\nu/5)f_0 \mu^2\big)^2 (1-8f_\nu) P_{\rm lin}(k) \ ,  
\label{eq:kaiser formula}
\end{equation}
where the relevant parameters controlling the amplitude of the power spectrum are then $b$ and $f_\nu$. 
Here $P_{\rm lin}$ is the linear power spectrum of matter as if it were all in the form of cold dark matter and baryons. 

Marginalising over $b$, from this naive Fisher we get $\widetilde{\sigma}_{f_\nu} = \sigma_{f_\nu} / \sqrt{1- \rho^2}$, 
where $\rho$ is the correlation coefficient between $b$ and $f_\nu$.  
The relative $\sigma$-sensitivity on $\Smnu$ is then given by $\sim f_\nu / \widetilde{\sigma}_{f_\nu} $. 
Here $\sigma_{f_\nu}$ and $\rho$ are obtained solving for
\begin{equation}
\begin{aligned}
\frac{1}{\sigma_{b}^2} \approx & \sum_{\ell,\ell'} \int_{\pmb k} \frac{\partial P_\ell(k)}{\partial {b}} \frac{1}{\sigma_{\ell\ell'}(k)^2}  \frac{\partial P_{\ell'}(k)}{\partial {b}} \ , \\
\frac{1}{\sigma_{f_\nu}^2} \approx & \sum_{\ell,\ell'} \int_{\pmb k} \frac{\partial P_\ell(k)}{\partial {f_\nu}} \frac{1}{\sigma_{\ell\ell'}(k)^2}  \frac{\partial P_{\ell'}(k)}{\partial {f_\nu}} \ , \\
\frac{\rho}{\sigma_{b} \sigma_{f_\nu}} \approx & \sum_{\ell,\ell'} \int_{\pmb k} \frac{\partial P_\ell(k)}{\partial {b}} \frac{1}{\sigma_{\ell\ell'}(k)^2}  \frac{\partial P_{\ell'}(k)}{\partial {f_\nu}} \ , 
\end{aligned}
\end{equation}
where assuming Gaussian errors with no shot noise, 
\begin{equation}
\sigma_{\ell \ell'}(k)^2 = \frac{(2\ell+1)(2\ell'+1)}{V_{\rm eff}} \int_{-1}^{1} d\mu \ P_g(k, \mu)^2 \mathcal{L}_\ell(\mu)\mathcal{L}_{\ell'}(\mu) \ .
\end{equation}
Since $\partial P_\ell(k) / \partial \theta_\alpha \propto P_{\rm lin}(k)$, with $\theta_\alpha = b, f_\nu$, and $\sigma_{\ell \ell'}(k) \propto P_{\rm lin}(k)$, the powers of $P_{\rm lin}(k)$ thus cancel in the Fisher elements above and we can then perform the integrals over $d^3k$, yielding $\int_{\pmb{k}} \ 1 \approx 4\pi \, k_{\rm max}^3 / (2\pi)^3 / 3$. 
At the end of the day, considering an effective $k_{\rm max} \sim 0.3 \ h/\unit{Mpc}$ (up to, roughly, where the signal gets shot-noise dominated), $V_{\rm eff} \sim 50 \unit{Gpc^3}/h^3$ from Table~\ref{tab:probes}, and fiducials $b \sim 2, f_0 \sim 0.68$, and $f_\nu \sim 0.4\%$ (corresponding to NO minimal mass), we get a relative $\sigma$-sensitivity of $f_\nu / \widetilde{\sigma}_{f_\nu} = \Smnu / \widetilde{\sigma}_{\Smnu} \sim \mathcal{O}(10)$, which matches well to the 1$\sigma$ sensitivity of 7 meV that we found for S4+MegaMapper in the NO ($\Smnu=60 \meV$). 

Thus, our rough analytical estimates (useful to cross-check the leading factors of $f_\nu$) fall in the ballpark of our refined Fisher analysis presented in the main text. 
Let us comment on that. 
In our analytic Fisher with only the power spectrum at tree level, $b$ and $f_\nu$ are highly correlated, i.e., $\rho= -0.99$. 
Therefore, any mean that can help in breaking this degeneracy can in principle drastically improve the bounds. 
In particular, nonlinearities (that are not considered in \cref{eq:kaiser formula}) help in breaking the correlation $\rho$ between $b$ and $f_\nu$: 
fixing all other nonlinear EFT parameters in the loop, we find $\rho = -0.95$, which naively would mean that the constraints would improve by a factor 2.
With the inclusion of the one-loop bispectrum, still keeping all EFT parameters fixed except $b$, we find that the degeneracy would reduce to $\rho = -0.7$ (which corresponds to an improvement of 5). 
In realistic settings, however, all EFT parameters are unknown and therefore, these additions come at the cost of a ``theoretical noise'' one needs to marginalise over. 
This nevertheless suggests that further information from higher-$N$ point functions or higher loops should improve the sensitivity on $\Smnu$.


\section{Fiducial model and parameter priors}
\label{sec:priors}

For cosmological parameters we impose a Gaussian prior on $\omega_{b}$ in the Fisher matrix motivated by Big-Bang Nucleosynthesis (BBN) experiments when analyzing LSS alone~\cite{Mossa:2020gjc}. When combining with CMB, this prior is not included. This is a prior with standard deviation $\sigma_{\rm BBN} = 0.00036$. This is the only prior imposed on cosmological parameters for the Fisher matrix, while for the CMB chains we use priors based on Planck~\cite{Planck:2018vyg}. Priors on beyond $\LCDM$ parameters $\Smnu$, $\log \Geff$, $\Neff$, $\delta m_{e}$,  and $\Ok$ are shown in~\cref{tab:priortab}.

\begin{table}[t]
\vspace*{1em}
\begin{tabular}{cccc}
\toprule
\multirow{2}{*}{\textbf{Parameter}} & \multicolumn{3}{c}{\textbf{Prior}}\\
&$\min$ & $\max$ &fiducial\\
\midrule
$\Smnu$ [eV] &  0 & 1.5 &0.06\\
$\log_{10} (\Geff^{\rm MI} \MeV^2)$ & $-5.5$ & $-3$ &-4.3\\
$\log_{10} (\Geff^{\rm SI}\MeV^2)$ & $-3$ & $-0.5$ &-1.8\\
$N_{\rm ur}\equiv \Neff-1$ & 0 & $\infty$ &2.0328\\
$\delta m_{e}$ & 0.1 & 10 &1.0\\
$\Ok$& $-0.5$ & 0.5 &0\\
$\ln 10^{10}A_{s}$& $-\infty$&$\infty$&3.044\\
$n_{s}$&$-\infty$&$\infty$&0.965\\
$h$&$-\infty$&$\infty$&0.673\\
$\omega_{\rm b}$&$-\infty$&$\infty$&0.02237\\
$\omega_{\rm cdm}$&$-\infty$&$\infty$&0.1203\\
\bottomrule
\end{tabular}
\caption{We report the minimum and maximum cosmological bounds used in MCMC analysis for Planck and S4 and fiducial cosmological background values used in Fisher forecast. The fiducial corresponds to $\theta_{\rm fid}$ for which the Fisher is evaluated at. In the case of $N_{\rm ur}$ no upper bound is specified.
}
\label{tab:priortab}
\end{table}

For the EFT parameters entering the Fisher matrix, we assign Gaussian priors following the same prescription as outlined in~\cite{DAmico:2022osl}. That is we assign Gaussian priors of width 2 on all parameters except for $c_{h,1}, c_{\pi,1}, c_{\pi v, 1}$ and $c^{\rm St}_{2}$, where we assign Gaussian priors of width 4. Furthermore, for the linear bias $b_{1}$, we choose to analyze this parameter in log space and utilize a log normal prior with standard deviation $\sigma_{\log b_{1}} = \sqrt{0.8}$. In addition, we have checked that loosening the prior widths of the EFT parameters by doubling the standard deviation $\sigma$ results in $\lesssim 8\%$ increase in neutrino bounds when analyzing Planck + DESI power spectrum only while including the bispectrum reduces this to $\lesssim 4\%$.

\PP{Furthermore, the Fisher forecast from the EFT likelihood is obtained imposing a perturbativity prior as described in~\cite{Braganca:2023pcp}.}
Roughly speaking the perturbativity prior is a theory prior imposing that the two-loop power spectrum and two-loop bispectrum as estimated from their one-loop scaling be bounded by the data noise.
This expectation on the size of the one-loop correction constrains the predictions within the physical region, in particular at low $k$'s where the data precision leaves more room. 
From the theoretical point of view, this is consistent prior to adopt. 
If we did not account for it, while the 1D marginalised constraints on $\Smnu$ in combination with CMB are degraded only up to $5\%$, the effect on the 2D correlation is more pronounced.


\section{Impact of individual neutrino masses}
\label{sec:m_nu_i}

\begin{figure}[h!]\centering
\begin{tikzpicture}
\node [above right,inner sep=0] (image) at (0,0) {
\includegraphics[width=.99\columnwidth]{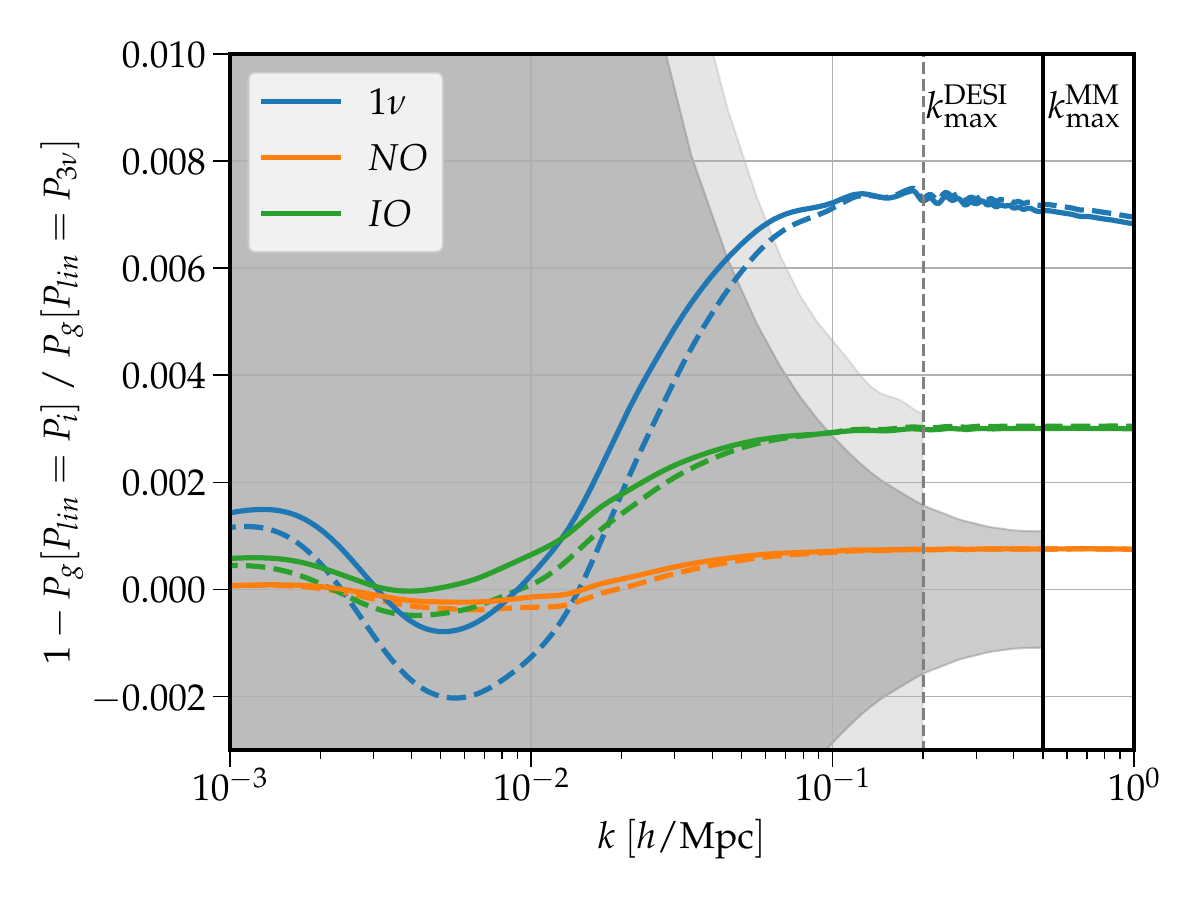}
};
\begin{scope}[x={($0.1*(image.south east)$)},
y={($0.1*(image.north west)$)}]
\node[fill=white,fill opacity=.7,text opacity=1,anchor=west] at (2.1,6.5) {$\Smnu = 100 \meV$};
\end{scope} 
\end{tikzpicture}
\caption{Relative differences in the galaxy power spectrum $P_g$ at fixed $\Smnu = 100 \meV$ for different assumptions about the individual neutrino masses. 
The spectra are normalised to the case of three massive neutrinos with degenerate masses. 
The coloured lines are computed with the linear power spectrum $P_\text{lin}$ obtained with only one massive neutrino (1$\nu$), normal (NO) or inverted ordering (IO), where either the fiducial redshift, linear bias, etc. for DESI (dashed lines) or MegaMapper (continuous lines) are used. 
The grey shaded areas represent the estimated error bars of DESI or MegaMapper, and the vertical lines show the estimated maximal $k$-reach in each experiment, $k_\textrm{max} \simeq 0.2 / 0.5 \hinvMpc$ respectively. 
To avoid clutter, we only show the monopole, and the data precision is a sum of the low and high redshift bins used in our Fisher forecast for bin size $\Delta k = 0.01 \hinvMpc$. 
}
\label{fig:pk_ordering_100meV}
\end{figure}

\begin{figure}[h!]\centering
\begin{tikzpicture}
\node [above right,inner sep=0] (image) at (0,0) {
\includegraphics[width=.99\columnwidth]{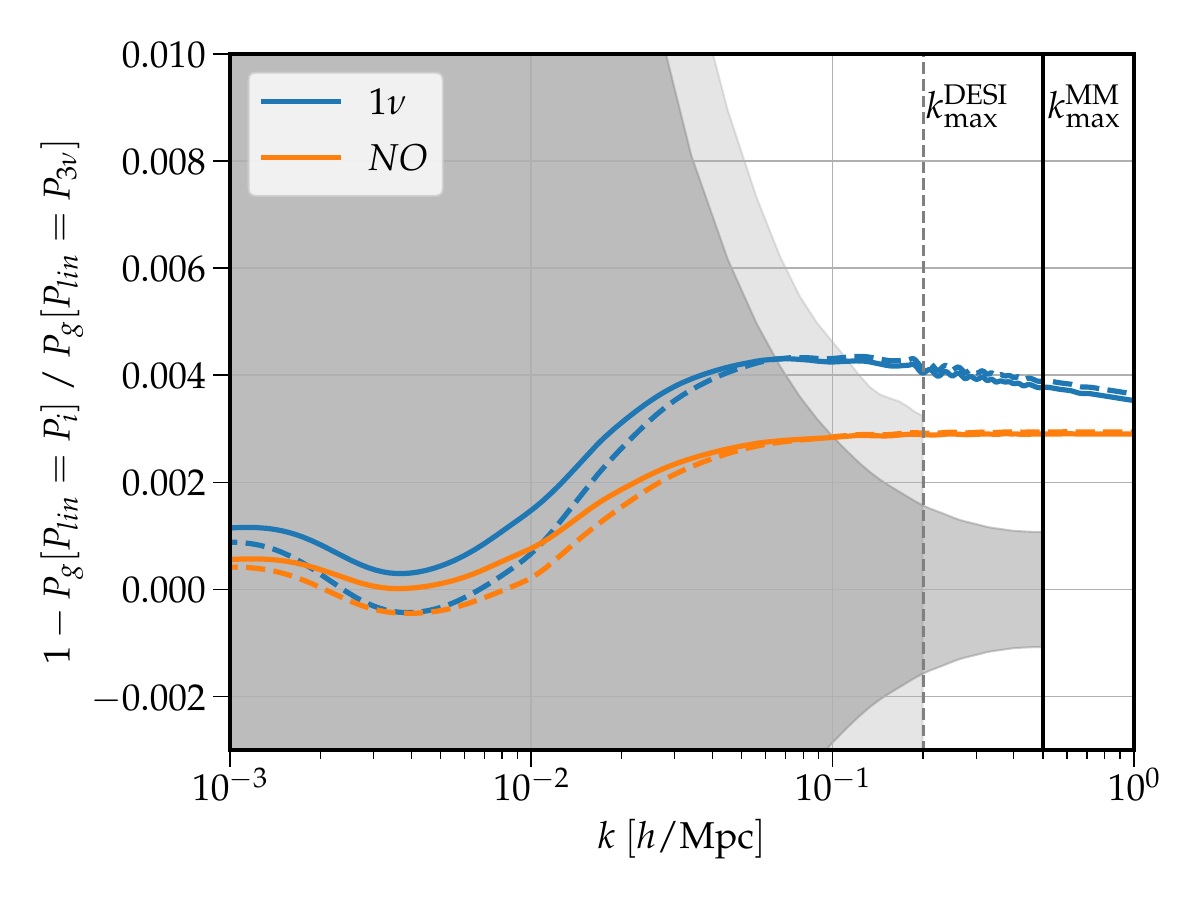}
};
\begin{scope}[x={($0.1*(image.south east)$)},
y={($0.1*(image.north west)$)}]
\node[fill=white,fill opacity=.7,text opacity=1,anchor=west] at (2.1,7.2) {$\Smnu = 60 \meV$};
\end{scope} 
\end{tikzpicture}
\caption{Same as \cref{fig:pk_ordering_100meV} but for $\Smnu=60\meV$. }
\label{fig:pk_ordering_60meV}
\end{figure}

In~\cref{fig:pk_ordering_100meV}~and~\cref{fig:pk_ordering_60meV}, we show the relative differences in the galaxy power spectrum at fixed $\Smnu$ for different assumptions about the individual neutrino masses. 

Compared to the case with three massive neutrinos with degenerate masses, we see that the usual so-called Planck prescription~\cite{Planck:2018vyg} with only one massive neutrino and two massless lead to the largest difference at $k\gg k_{\rm FS}$, of about $0.7\%$ and $0.4\%$ for $\Smnu = 100 \meV$ and $\Smnu = 60 \meV$, respectively. 
The differences to the cases assuming either normal or inverted ordering are smaller, of at most $\sim 0.3\%$, and the two ordering lead to a difference between themselves of $\sim 0.2\%$ for $\Smnu = 100 \meV$. 
Compared to the relative data precision at $k\gg k_{\rm FS}$ of about $0.3\%$ for DESI and $0.1\%$ for MegaMapper, these differences appear significant. 
However, because most of the scale-dependent features are at $k\lesssim k_{\rm FS}$ where the data precision is instead loose, one can anticipate that the mostly flat differences visible at high-$k$ can be easily absorbed in a rescaling of the amplitude, e.g., $b_1$. 
We can draw two conclusions. 
First, this justifies the use of approximate three degenerate neutrino masses for cosmology~\cite{Archidiacono:2020dvx,Scott:2024rwc}, while considering only one massive neutrino appears to be reasonable for $\Smnu = 60 \meV$~\cite{Planck:2018vyg}. 
Second, we do not expect in practice that the forecast experiments will be sensitive enough to the mass ordering. 
We do not expect the addition of the bispectrum of galaxies to change these conclusions. 
In light of this discussion, the synergy between neutrino oscillation experiments and cosmology appears to be crucial to pin point the details about the neutrino sector in the forthcoming years.


\bibliography{bib_mnu-LSS.bib}

\end{document}